\documentclass[preprint,12pt]{elsart}
\usepackage{latexsym}
\usepackage{amssymb}
\usepackage{graphicx}
\usepackage{epsfig}
\usepackage{wasysym}
\usepackage{esint}
\usepackage{axodraw}

\newcommand{\beq}[0]{\begin{equation}}
\newcommand{\eeq}[0]{\end{equation}}

\newcommand{\bit}{\begin{itemize}}
\newcommand{\eit}{\end{itemize}}
\newcommand{\bea}{\begin{eqnarray*}}
\newcommand{\eea}{\end{eqnarray*}}
\newcommand{\beanmb}{\begin{eqnarray}}
\newcommand{\eeanmb}{\end{eqnarray}}
\newcommand{\ben}{\begin{equation}}
\newcommand{\een}{\end{equation}}
\newcommand{\dr}{{\mathrm d}}
\newcommand{\er}{{\mathrm e}}

  \newcommand{\ccaption}[2]{
    \begin{center}
    \parbox{0.85\textwidth}{
      \caption[#1]{\small{{#2}}}
      }
    \end{center}
    }

\begin{document}
\newlength{\caheight}
\setlength{\caheight}{12pt}
\multiply\caheight by 7
\newlength{\secondpar}
\setlength{\secondpar}{\hsize}
\divide\secondpar by 3
\newlength{\firstpar}
\setlength{\firstpar}{\secondpar}
\multiply\firstpar by 2

\begin{frontmatter}
\vskip 48pt
\title{A Fully Numerical Approach to One-Loop Amplitudes}
\author[1]{M.~Moretti}
\author[2]{F.~Piccinini}
\author[3]{A.D.~Polosa}
\address[1]{Dipartimento di Fisica, Universit\`a di Ferrara\\
Via Saragat 1, I-44100, Ferrara, Italy}
\address[2]{Istituto Nazionale di Fisica Nucleare, Sezione di Pavia\\
Via A. Bassi 6, 27100, Pavia, Italy}
\address[3]{Istituto Nazionale di Fisica Nucleare, Sezione di Roma\\
P.le A. Moro 2, I-00185, Roma, Italy}
\begin{abstract}
We suggest a new approach for the automatic and fully 
numerical evaluation of one-loop scattering amplitudes 
in  perturbative quantum field theory. 
We use  suitably formulated dispersion relations to 
perform the calculation as a convolution of
tree-level amplitudes. This allows to take advantage of 
the iterative numerical algorithms for the evaluation 
of leading order matrix elements. 
\end{abstract}

\end{frontmatter}
\newpage
\section{Introduction}
\label{intro}
The LHC   will provide unprecedented experimental and theoretical 
challanges. Among others, the presence of many final state jets 
for any interesting observable, makes absolutely crucial a better control of 
theoretical predictions and in particular of higher order contributions. 
At present several Monte Carlo event 
generators based on exact Leading Order (LO) matrix elements 
are available such as  
{\tt Alpgen}~\cite{alpgen}, {\tt Helac}~\cite{helac}, 
{\tt MadEvent}~\cite{madevent} 
and {\tt Sherpa}~\cite{sherpa}. These codes are of extreme importance 
for the modeling of multi-jet final states. 
Nevertheless, the main problem connected with LO calculations 
is the strong factorisation/renormalisation scale dependence. 
In this respect it would be useful to have also Next-to-Leading Order 
(NLO) calculations for such multi-jet final states in order to better control 
the overall normalization of the theoretical predictions. 
Given the huge complexity of the calculations, 
many theoretical efforts have been recently 
 devoted to the development of new approaches to the problem. 

The real part of the NLO corrections of 
any $n$-body final state, given by  $(n+1)$-body tree-level matrix elements, 
can be computed efficiently thanks to automated 
helicity amplitudes~\cite{helicity}
or fully numerical algorithms~\cite{alpha,alphaqcd}, 
which allowed the development of the above mentioned LO matrix element 
event generators. 
On the other hand the NLO virtual corrections for a generic 
$2 \to n$ process are 
known for $n=2$ and only for some $2 \to 3$ processes. Among the latter 
it is worth mentioning the QCD corrections to $pp \to ZZZ$, $WWZ$, 
$HHH$~\cite{vvv}, $pp \to H j j$~\cite{hjj} and 
$pp \to t {\bar t} j$~\cite{ttj}. For $2 \to 4$ processes, only 
the QCD corrections to the weak boson fusion 
$p p \to W W j j$~\cite{vbf1}, $WZjj$~\cite{vbf2} and 
the complete electroweak ${\mathcal O}(\alpha)$ corrections 
to $e^+ e^- \to 4$~fermions~\cite{4f} have been calculated 
with standard diagrammatic approaches, supplemented with 
new reduction techniques of scalar and tensor integrals~\cite{reduct}. 
No complete calculation of NLO QCD corrections to $p p \to 4$~partons 
exists at present. Only the calculation of 
six-gluon scattering amplitudes has been completed successfully~\cite{6g}, 
even if the needed CPU time renders the calculation unsuited 
for event generation. Within QED, the calculation of $n-$photon 
amplitudes have been performed by several groups, with $n$ up to 
six~\cite{6gamma}.

In order to overcome the difficulties in 
the calculation of virtual corrections to multi-leg processes, several 
groups have developed new (semi)numerical and analytical techniques. 
It is worth mentioning the work of Ref.~\cite{fppu}, 
based on the method developed in Ref.~\cite{gp}, which has been 
used to build up the complete two-loop renormalization program 
within the Standard Model~\cite{2-loop}. 
Pure numerical algorithms have been proposed 
in Ref.~\cite{ds}, with applications to $e^+ e^-$ annihilations. 

The analytical methods exploit the unitarity cut method and on-shell recursion 
relations to calculate the so-called cut-constructible part of loop 
amplitudes (see Ref.~\cite{recursion} 
for an exhaustive review and references). The method is particularly 
suited for loop calculations in gauge theories, where the 
final analytical expressions are quite simple. 
An efficient algebraic technique, suitable for numerical implementation, 
has been suggested in Ref.~\cite{opp}. 
The main problem of the 
unitarity cut approach is related to the so-called rational 
part of the amplitude, for which other methods are 
required~\cite{rational}. 
Very recently, in Ref.~\cite{gkm} it has been proposed 
an algorithm based on $D$-dimensional unitarity, which 
allows the calculation of both cut-constructible and rational parts 
of one-loop scattering amplitudes. Moreover the method reduces 
the algorithmic complexity at the polynomial level and it is 
suited for numerical implementation. 

The common feature of the above mentioned techniques is the use 
of dimensional regularization and the decomposition of the 
amplitude in linear combinations of basic scalar loop integrals. 

In the present note we  study an alternative way, of purely 
numerical nature, for the calculation of loop-amplitudes, 
exploiting the fact that we have already at hand efficient and 
well developed numerical 
tools for the calculation of tree-level scattering amplitudes.
We follow the approach suggested by Veltman in Ref.~\cite{veltman71}, 
according to which any one-loop amplitude can be written as a convolution 
of LO amplitudes by means of appropriate  dispersion relations. The method 
has been used in Refs.~\cite{veltman71,diagrammar} at the formal level 
to discuss causality for individual diagrams. 
With a particular way of balancing energy conservation in cut diagrams
by adding an `artificial particle', a Feynman graph can 
be computed by the convolution of LO amplitudes, 
cutting any two internal lines. The method 
can be reformulated in terms of amplitudes, rather than in terms 
of individual Feynman graphs, and in this way 
we can retain the main advantage of LO approach, namely the power law
growth of computational complexity rather than factorial 
growth. 

The numerical nature of the procedure is an added 
bonus for the automation of the procedure. 
In the calculation of a scattering cross sections, 
the  loop integration is performed through the Monte Carlo 
method together with the phase space integration. This leaves 
two important open problems: the stability of the numerical 
integration and the regularization method for infrared and collinear 
singularities, which needs to be performed in four dimensions.

In the following sections we illustrate the method for the simple 
case of the $\phi^3$ model, showing some numerical comparisons 
between the results obtained with this approach and the 
standard diagrammatic techniques. 

\section{Introduction to the method}

We shall first shortly review the largest time equation and the dispersion
relations ``\`a la Veltman'' we shall use in this paper, referring to
\cite{veltman71} for furthers\footnote{We follow closely 
ref.~\cite{veltman71},
using Minkowski metrics and this justify some sign difference and
the appearence of constant factors $i$. We also 
use the relationship $\Delta_{xy}^+ = \Delta_{yx}^-$ everywhere in the paper.
}
 details.

For definitness we work in a $\phi^3$ theory,
setting the coupling constant to one for simplicity. The propagator
 for
a scalar field, propagating from spacetime point $x$ to $y$, is
\beanmb
\Delta_{xy} & = &
\int \frac {\dr l }{(2 \pi)^4} \er^{-i l (x-y)}\frac{i}{l^2-m^2} =
  \theta(x_0-y_0) \Delta_{xy}^+  + \theta(y_0-x_0) \Delta_{yx}^+
 \label {prpg}
 \\
\Delta_{xy}^+ & = &
\int \frac {\dr l }{(2 \pi)^3} \er^{-i l (x-y)} \theta(l_0) \delta (l^2-m^2)
\label{dltpls}
\eeanmb
For a generic n-point, one-particle irreducible, Feynman diagram $G_n$
we obtain, in configuration space:
\beq
G_n(x_1,\dots x_n) =  i^n \Delta_{x_1 x_2} \Delta_{x_2 x_3} \dots \Delta_{x_n x_1} 
\label{eq:npnt}
\eeq
where $ x_j$ enters clockwise in the diagram (see fig.~\ref{npoint}).
We now introduce a new notation: the "underlined configuration variable"
$ \underline x_j$ and a new set of functions
\beanmb
\tilde G_n(x_1,\dots, \underline
x_j,\dots, \underline x_k, \dots x_n) = &&  \nonumber \\
(-1)^{n_u}
i^n \Delta_{x_1 x_2} 
 \dots &&
\Delta_{x_{j-1} \underline x_j}
 \dots 
\Delta_{x_k \underline x_{k+1}}
\dots
\Delta_{x_n x_1}    
\label{eq:npntund}
\eeanmb
where $n_u$ is the number of underlyined variables and
\bea
\Delta_{\underline x_l x_m} & = &  \Delta^+_{ x_m x_l} \\
\Delta_{x_l \underline  x_m} & = &  \Delta^+_{ x_l x_m} \\
\Delta_{\underline x_l \underline x_m} & = &
\Delta^*_{ x_l x_m}
\eea
Notice that because
of the $\theta(l_0) $ term in equation
(\ref{dltpls}) {\em energy is alway required to flow towards the
underlined vertex}. 

\newpage

\begin{center}
\begin{figure}[hbt]
\SetScale{0.5}
\SetWidth{1.}
{\unitlength 0.5pt
\begin{picture}(0,0)(-50,50)
\Line(180,0)(460,0)
\Line(220,0)(220,-200)
\Line(420,0)(420,-200)
\Line(180,-200)(460,-200)
\put(190,-30){$x_1$}
\put(450,-30){$x_2$}
\put(190,-230){$x_4$}
\put(450,-230){$x_3$}
\end{picture} 
}
\vskip 150pt
\ccaption{}{\label{npoint}
Four point function
$G_4(x_1,x_2,x_3,x_4)$.
}
\end{figure}
\end{center}

We can introduce a diagrammatic notation
for the $\tilde G_n$ functions: we shall denote with a small
circle a vertex to which an underlined variable is
``attached'', namely, for example:
\begin{center}
\begin{figure}[hbt]
\SetScale{0.4}
\SetWidth{1.}
{\unitlength 0.4pt
\begin{picture}(0,0)(-150,50)
\GCirc(220,0){8}{1}
\GCirc(320,-100){8}{1}
\Line(220,0)(420,0)
\Line(220,0)(320,-100)
\Line(420,0)(320,-100)
\put(190,-10){$x$}
\put(430,-10){$y$}
\put(290,-110){$z$}
\begin{Large}
\put(-50,-70){$\tilde G(\underline x, y, \underline z) =$ }
\end{Large}
\end{picture} }
\vskip 100pt
\end{figure}
\end{center}
\vskip -50pt
Therefore the set of Feynman rules for an
underlined diagram are:
\begin{center}
\SetScale{0.4}
\SetWidth{1.}
\unitlength 0.4pt
\begin{figure}[hbt]
\vskip 12pt
\begin{picture}(0,0)(-30,20)
\Line(100,0)(130,0)
\Line(130,0)(150,20)
\Line(130,0)(150,-20)
\begin{Large}
\put(170,-10){$\equiv i $; }
\end{Large}
\end{picture} 
\begin{picture}(0,0)(-340,20)
\GCirc(130,0){8}{1}
\Line(100,0)(130,0)
\Line(130,0)(150,20)
\Line(130,0)(150,-20)
\begin{Large}
\put(170,-10){$\equiv - i $ }
\end{Large}
\end{picture}
\begin{picture}(0,0)(0,130)
\put(70,-10){$x$}
\put(150,-10){$y$}
\Line(100,0)(140,0)
\begin{Large}
\put(180,-10){$\equiv \Delta_{xy}; $ }
\end{Large}
\end{picture}
\begin{picture}(0,0)(-300,130)
\put(65,-10){$x$}
\put(160,-10){$y$}
\GCirc(140,0){8}{1}
\GCirc(100,0){8}{1}
\Line(100,0)(140,0)
\begin{Large}
\put(190,-10){$\equiv \Delta^*_{xy}; $ }
\end{Large}
\end{picture}
\begin{picture}(0,0)(0,240)
\GCirc(115,0){8}{1}
\put(50,-10){$x$}
\put(130,-10){$y$}
\Line(80,0)(115,0)
\begin{Large}
\put(160,-10){$\equiv \Delta^+_{xy}; $ }
\end{Large}
\end{picture}
\begin{picture}(0,0)(-300,240)
\GCirc(80,0){8}{1}
\put(45,-10){$x$}
\put(135,-10){$y$}
\Line(80,0)(120,0)
\begin{Large}
\put(170,-10){$\equiv  \Delta^+_{yx} $ }
\end{Large}
\end{picture}
\end{figure}
\vskip 100pt
\end{center}
\vskip -12pt
With the above set of Feynman rules and definitions
we obtain
\beq
x_0 > y_0,z_0 \Rightarrow \tilde G(x,y,z)=
-\tilde G(\underline x,y,z); \quad \tilde G(x,\underline y,z)=
-\tilde G(\underline x,\underline y,z); \dots
\label{eq:paircanc}
\eeq
namely if $x_0$ is the largest time ``entering into a given
graph'' the sum of a $\tilde G$ with underlined $x$ and
 without underlined $x$ is equal to zero. This feature
applies to any diagram with an arbitrary number of legs.
This observation leads to the largest time equation
\beq
\sum_{\mathrm all \ possible \ underlings} 
\tilde G(x_1, \dots , x_n) = 0 
\label{eq:lrgst}
\eeq
which, for the case of a three point function,
reads
\beanmb
& & 
\tilde G(x,y,z)+ \tilde G(\underline x,y,z) + 
\tilde G(x,\underline y,z) + \tilde G(x,y,\underline z) 
+ \tilde G(\underline x,\underline y,z)
 + \tilde G(\underline x,y,\underline z)  
\nonumber \\ 
& & 
+ \tilde G(x,\underline y,\underline z)  
+ \tilde G(\underline x,\underline y,\underline z) = 0
\label{eq:lgst3pt}
\eeanmb
Indeed if $x_0 > y_0,z_0$ the contributions
(1,2), (3,5), (4,6) and (7,8) of Eqn.~(\ref{eq:lgst3pt})
cancel pairwise. If $y_0 > x_0,z_0$ the cancellation
occurs among the (1,3), (2,5), (4,7) and (6,8) pair 
of diagrams. Finally if $z_0 > x_0,y_0$ the cancellation
occurs among the (1,4), (2,6), (3,7) and (4,8) pair 
of diagrams.
This cancellation is explicitly worked out in the appendix.
From Eqn.~(\ref{eq:lgst3pt}) one easily arrive at the 
standard Cutkowsky rules for the imaginary\footnote{Notice
that, using the conventional decomposition of the S matrix
$S= I + i T$, $G(x_1,\dots,x_n)$ contributes to $i T$} part 
of a diagram
\beanmb
\tilde G(x,y,z) + \tilde G(\underline x,\underline y,\underline z)  & = &
G(x,y,z) +G(x,y,z)^* \nonumber \\ 
& = & - \left [
\tilde G(\underline x,y,z) + 
\tilde G(x,\underline y,z) + \tilde G(x,y,\underline z) 
+ \tilde G(\underline x,\underline y,z) \right .
 \nonumber \\ 
&  &
\left . + \tilde G(\underline x,y,\underline z)  
+ \tilde G(x,\underline y,\underline z)
\right ] \label {eq:unitarity} \\ \nonumber 
\eeanmb
Since the Feynman rule for a line connecting an ordinary
 vertex and an underlined one is indeed equivalent
to cut the diagram on the corresponding line
this is the standard Cutkowsky rule
for the imaginary part of the diagram. Notice that several
contributions in the above equation vanish,
since
energy is required to flow from standard to underlined 
vertices and conflicting conditions might be required
to hold simultaneously.
By mean of the largest time equation we can derive 
another equation which allows to formulate a universal
dispersive like formula for an arbitrary Feynman graph.
We choose any\footnote{that's why we choose to discuss
both the simpler 
two point function and the three point one: 
for the two point function the arbitrairness in the choice
of the two vertices has no meaning.}
 two vertices in the diagram of interest,
say $y$ and $z$ in the three point function.
We have 
\beq
G(x,y,z) = \theta(y_0-z_0) G(x,y,z) + 
\theta(z_0-y_0) G(x,y,z)
\label{eq:thetasum}
\eeq
and
\beq
\theta(y_0-z_0) \left [
 \tilde G( x,y, z) +
  \tilde G(x, \underline y,  z)
+ \tilde G(  \underline x, y,  z) 
+  \tilde G(\underline x,\underline y,  z)
\right ] = 0 
\label{eq:thetacanc}
\eeq
Indeed, as discussed in Eqn.~(\ref{eq:paircanc}),
if $y_0 > z_0,x_0$ contributions (1,2) and (3,4) cancel
pairwise; if $x_0 > y_0,z_0$ 
 contributions (1,3) and (2,4) cancel
pairwise; if $z_0 > x_0,y_0$ the overall
$\theta$ function ensures the vanishing of this contribution.
From Eqns.~(\ref{eq:lgst3pt}),
 (\ref{eq:thetasum}) and (\ref{eq:thetacanc})
(and the analogous one for the term with $\theta(z_0-y_0)$
we obtain
\beanmb
G(x,y,z) & = & -\theta(y_0-z_0) \left [ \tilde 
G(\underline x,y, z) + \tilde 
G( x,\underline y, z) + \tilde 
G(\underline x,\underline y, z)
\right ] \nonumber \\ & & 
-\theta(z_0-y_0) \left [ \tilde 
G(\underline x,y, z) + \tilde 
G( x,y, \underline z) + \tilde 
G(\underline x, y, \underline z)
\right ]
\nonumber \\
& = &
-\theta(y_0-z_0) \left [ \tilde 
G( x,\underline y, z) + \tilde 
G(\underline x,\underline y, z)
\right ] \nonumber \\ 
& & -\theta(z_0-y_0) \left [  \tilde 
G( x,y, \underline z) + \tilde 
G(\underline x, y, \underline z)
\right ]
 - \tilde 
G( {\underline x},y, z) 
\label{eq:causality}
\eeanmb
Notice that the last term in Eqn.~(\ref{eq:causality})
corresponds to a cut in which the two chosen vertices
$y$ and $z$ lie on the same side of the cut. This term
is one of the terms occurring in the unitarity
relation (\ref{eq:unitarity}) and is purely real.

For twice 
the imaginary part of $G$ (real part of the amplitude),
from Eqn.(\ref{eq:causality}), we obtain
\beanmb
G(x,y,z) - G(x,y,z)^* & = & 
[ \theta(y_0-z_0)-\theta(z_0-y_0) ] \left [  \tilde 
G( x,\underline y,  z) +  \tilde 
G( \underline x, \underline y,  z) \right .
\nonumber \\ & & 
\left .
-\tilde 
G( x,\underline y, z) - \tilde 
G(\underline x,\underline y, z) \right ]
\label{eq:realpart3pt}
\eeanmb
For the two point function we obtain 
\beanmb
G(x,y) - G(x,y)^* & = & [\theta(x_0-y_0)-\theta(y_0-x_0) ] 
\left [  \tilde 
G( x,\underline y) + \tilde G( y,\underline x) \right ]
\label{eq:realpart2pt} \\
G(x,y) + G(x,y)^* & = & 
\left [  \tilde 
G( x,\underline y) + \tilde G( y,\underline x) \right ]
\label{eq:imagpart2pt}
\eeanmb

This is a dispersion relation for the real
part of a generic, one particle irreducible
n-point function.
In the appendix we verify explicitly  that
 Eqn.~(\ref{eq:realpart2pt}) holds and we show that
inserting the integral representation of the
$\theta$ function it leads to a dispersive like integral.

We introduce a diagrammatic representation for
the $\theta$ function appearing in 
Eqns.~(\ref{eq:realpart3pt}) and (\ref{eq:realpart2pt}): 
we will connect the two chosen vertices with a wavy line
herafter referred to as the $\tau$-line.
Before providing the generic recipe, notice
that a function $\tilde G$ is non vanishing only
if ordinary and underlined vertices belong to two
distinct connected regions:
the one containing underlined variables
will be referred to as shadowed the other one
as unshadowed. The energy along the cut lines will
flow from the unshaded to the shaded region.
We refer to 
\cite{veltman71} for a thorough discussion
and we just give an example of a vanishing contribution
to a box function in fig.~\ref{box0}.
\begin{center}
\begin{figure}[hbt]
\SetScale{0.5}
\SetWidth{1.}
{\unitlength 0.5pt
\begin{picture}(0,0)(-50,50)
\ArrowLine(180,0)(220,0)
\ArrowLine(180,-200)(220,-200)
\ArrowLine(420,0)(460,0)
\ArrowLine(420,-200)(460,-200)
\ArrowLine(420,0)(220,0)
\ArrowLine(220,-200)(220,0)
\ArrowLine(420,0)(420,-200)
\ArrowLine(220,-200)(420,-200)
\GCirc(220,0){8}{1}
\GCirc(420,-200){8}{1}
\put(190,-30){$P_1$}
\put(450,-30){$P_2$}
\put(190,-230){$P_4$}
\put(450,-230){$P_3$}
\GlueArc(320,-80)(130,40,140){5}{8.57}
\end{picture} 
}
\vskip 150pt
\ccaption{}{\label{box0}
Vanishing contribution to the
four point function. 
The diagram is split into four disconnected regions (each propagator
connecting a dotted and an undotted vertex is on shell).
Arrows denote the direction of the energy flow
}
\end{figure}
\end{center}

For a generic Green function the method can be applied 
as well and, summarizing the results sketched above,
leads to the following modified
Feynman rules:
\begin{itemize}
\item Draw a $\tau-$line (the $\theta$ function insertion 
in Eqn.~(\ref{eq:thetasum}))
between two arbitrary vertices of the Feynman graph.
\item Cut the graph in two disconnected parts in such a way that the
$\tau-$line crosses 
(if both the chosen vertices are on the same side of the
cut this will contribute only to the imaginary part
of the graph as $ \tilde 
G(\underline x,y, z) $ in Eqn.(\ref{eq:causality}))
the cut.
\item In the vertices where the  $\tau-$line is 
absorbed/emitted it does contribute (this and the
following items are 
esplicitly shown,
for the two point function, in the appendix)
to the energy balance with a $\tau$.
\item The $\tau-$line contributes a factor 
$1/(i\pi)\times \fint d\tau \,  ( 1/\tau )$ ($\fint$ denotes 
the principal value integral).
\item Cut internal lines contribute a factor $\theta(\pm k^0)\delta(k^2-m^2)$ where $\pm k^0$ is chosen in such a way that their energy flows towards the shaded region.
\item A two-body phase space integral is associated with the two cut internal lines:
$d\Phi=\;\lambda^{1/2}(P^2,p_1^2,p_2^2)/P^2 \; d\cos\theta d\phi$ where $p_{1,2}$ are the momenta of the cut internal lines, and $P$ is the sum of the external 4-momenta $k_j$ which enter into the unshaded part of the graph plus the $\tau-$line contribution which amounts to subtract $\tau$ to the time-like $P$ component.
\item Sum over the contributions of all allowed cuts (see the items above).
\end{itemize}

\subsection{The two point function}
\label{2points_sub}
In order to illustrate the method let us start by computing the simplest 1-loop diagram in a scalar $\phi^3$ theory ($q$ is off-shell):
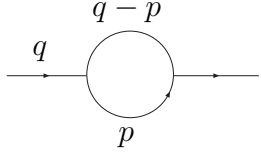
\begin{center}
\SetScale{0.5}
\SetWidth{1.}
\unitlength 0.5pt
  \begin{picture}(418,75) (30,-106)
    \SetWidth{0.5}
    \SetColor{Black}
    \put(140,-135){$q$}
    \put(185,-105){$q-p$}
     \put(205,-200){$p$}
      
    \ArrowLine(120,-150)(180,-150)
    \ArrowArc(213,-149)(32.65,153,513)
    \ArrowLine(246,-150)(310,-150)
        \end{picture}
\end{center}
\vskip1.5truecm
Using the standard techniques and omitting all constant factors for the sake of simplicity, 
this loop diagram is proportional to:
\begin{equation}
\int d^4p \frac{1}{(p^2-m^2+i\epsilon)((q-p)^2+i\epsilon)}=\int_0^1 dx
\int d^4p\frac{1}{(p^2+L(x))^2}
\label{loopi}
\end{equation}
where one of the fields in the loop is assumed to be massless and
\begin{equation}
L(x)=x(1-x)q^2-(1-x)m^2.
\end{equation}
Upon rotation to euclidean momenta and setting an ultraviolet cut-off $\Lambda$,
the loop integral in (\ref{loopi}) reduces to
\begin{eqnarray}
&&\pi^2 i\int_0^1 dx\int_0^{\Lambda^2}dz\frac{z}{(z-L(x))^2}\simeq\nonumber\\
&&\simeq  \pi^2 i\int_0^1 dx [\ln \Lambda^2-\ln(-x(1-x)q^2+(1-x)m^2)]=\nonumber\\
&&=\pi^2 i \left(\ln\Lambda^2-\frac{q^2-m^2}{q^2}\ln |q^2-m^2|-\frac{m^2}{q^2}\ln m^2+2 \right. \nonumber \\
&& \left. \quad \quad \quad +i\pi 
\frac{q^2-m^2}{q^2}\theta(q^2-m^2)\right)\nonumber
\end{eqnarray}
where in the latter step we have used integration by parts, and we have also used 
$\Lambda^2\gg q^2,m^2$.

The largest time  equation (\ref{eq:imagpart2pt}), 
for the two
poin function, reads
\bea
 G(x,y)+G(x,y)^* & = & (\Delta^+_{xy})^2 + (\Delta^+_{yx})^2
\nonumber \\
G(q)+G(q)^*& = & -2\pi^3 \frac{q^2-m^2}{q^2}\theta(q^2-m^2)
\label{result}
\eea
$G(q)$ denoting the fourier transform of $G(x,y)$.

In fact, we have:
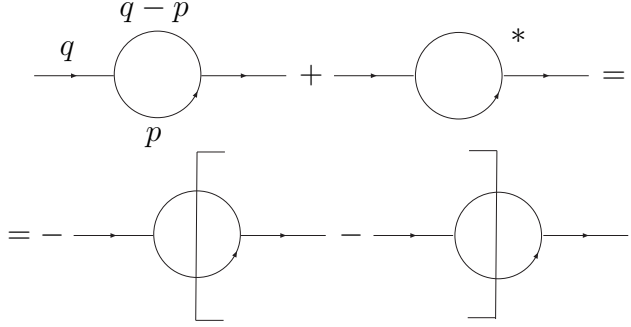
\begin{center}
\SetScale{0.5}
\SetWidth{1.}
\unitlength 0.5pt
  \begin{picture}(418,75) (150,-106)
    \SetWidth{0.5}
    \SetColor{Black}
    \put(140,-135){$q$}
    \put(321,-156){$+$}
    \put(481,-126){$*$}
    \put(185,-105){$q-p$}
     \put(205,-200){$p$}
      \put(550,-156){$=$}
    \ArrowLine(120,-150)(180,-150)
    \ArrowArc(213,-149)(32.65,153,513)
    \ArrowLine(246,-150)(310,-150)
    \ArrowLine(346,-150)(406,-150)
    \ArrowArc(440,-150)(32.65,153,513)
    \ArrowLine(474,-150)(538,-150)
    \end{picture}  
\vskip1.3truecm
\begin{picture}(418,128) (120,-161)
    \SetWidth{0.5}
    \SetColor{Black}
    \ArrowLine(120,-97)(180,-97)
    \ArrowArc(213,-96)(32.65,153,513)
    \ArrowLine(246,-97)(310,-97)
    \ArrowLine(346,-97)(406,-97)
    \ArrowArc(440,-97)(32.65,153,513)
    \ArrowLine(474,-97)(538,-97)
    \Line(213,-34)(212,-161)
    \Line(439,-33)(438,-160)
    \Line(213,-34)(234,-34)
    \Line(212,-161)(233,-161)
    \Line(418,-33)(439,-33)
    \Line(417,-159)(438,-159)
   \put(321,-103){$-$}
   \put(70,-103){$=-$}
  \end{picture}
    \end{center}
\vskip0.5truecm

where the parenthesis `$[  $' indicates that the shadowed part of the diagram is on the right and vice-versa for `$]$'. The rule for internal lines flowing towards the shaded region is to associate to them a $\theta(k^0)\delta(k^2-m)$. For a line flowing out of the shaded region one associates a 
 $\theta(-k^0)\delta(k^2-m)$. As a result, the latter diagrammatic equation reduces to (apart from an overall sign)
 \begin{eqnarray}
 &&\int d^4 p [\theta(-q^0+p^0)\delta((q-p)^2)\theta(p^0)\delta(p^2-m^2)+\nonumber\\
 &&\;\;\;\;\;\;\;\; +\theta(q^0-p^0)\delta((q-p)^2)\theta(-p^0)\delta(p^2-m^2)]
\end{eqnarray}
The first term of the latter expression can be written introducing a Dirac-$\delta$ as
\begin{equation}
\int d^4 p\int d^4 Q \;\delta^4(q-p-Q) \theta(Q^0)\delta(Q^2)\theta(p^0)\delta(p^2-m^2).
\end{equation}
which simply reduces to the phase space factor:
\begin{equation}
\int \frac{d^3p}{2p^0}\frac{d^3Q}{2Q^0}\delta^4(q-p-Q)=\frac{\pi}{2}\frac{\lambda^{1/2}(q^2,m^2,0)}{q^2}=\frac{\pi}{2}\frac{q^2-m^2}{q^2}\theta(q^2-m^2),
\end{equation}
where the $\theta$ factor is the threshold condition $q^2>m^2$. For simplicity, we have neglected constant factors. Computing also the second integral and accounting for the proper factors of $\pi$,
one finds (see eq.~(\ref{result})):
\begin{equation} 
F+F^*=-2\pi^3\frac{q^2-m^2}{q^2}\theta(q^2-m^2)
\label{rlslf}
\end{equation}

Let us now turn to the remaining part of our loop diagram, namely the one obtainable by subtracting 
$F-F^*$. For this purpose, following~\cite{veltman71} we consider the diagrammatic equation:
\begin{center}
\SetScale{0.5}
\SetWidth{1.}
\unitlength 0.5pt
  \begin{picture}(418,75) (150,-106)
    \SetWidth{0.5}
    \SetColor{Black}
    \put(140,-135){$q$}
    \put(321,-156){$-$}
    \put(481,-126){$*$}
    \put(185,-105){$q-p$}
     \put(205,-200){$p$}
      \put(550,-156){$=$}
    \ArrowLine(120,-150)(180,-150)
    \ArrowArc(213,-149)(32.65,153,513)
    \ArrowLine(246,-150)(310,-150)
    \ArrowLine(346,-150)(406,-150)
    \ArrowArc(440,-150)(32.65,153,513)
    \ArrowLine(474,-150)(538,-150)
    \end{picture}  
\vskip1.3truecm    
    \begin{picture}(418,140) (145,-161)
    \SetWidth{0.5}
    \SetColor{Black}
    \ArrowLine(120,-85)(180,-85)
    \ArrowArc(213,-84)(32.65,153,513)
    \ArrowLine(246,-85)(310,-85)
    \ArrowLine(346,-85)(406,-85)
    \ArrowArc(440,-85)(32.65,153,513)
    \ArrowLine(474,-85)(538,-85)
    \Line(213,-22)(212,-149)
    \Line(439,-21)(438,-148)
    \Line(213,-22)(234,-22)
    \Line(212,-149)(233,-149)
    \Line(418,-21)(439,-21)
    \Line(417,-147)(438,-147)
    \GlueArc(213,-100.50)(33.45,386.67,153.33){-5}{8.57}
    \GlueArc(440,-100.50)(33.45,386.67,153.33){-5}{8.57}
    \LongArrowArc(442.5,-97.79)(59.39,-130.41,-49.59)
    \LongArrowArcn(212.5,-97.79)(59.39,-49.59,-130.41)
    \put(100,-90){$=$}
    \put(320,-90){$+$}
  \end{picture}
\end{center}
\vskip1truecm
Here we have a new symbol, the so called $\tau$-line represented 
by a $\gluon$. The $\tau$-line makes energy flow from the 
unshadowed to the shadowed areas~\footnote{
More precisely the $\tau$-line ensures: 1) that the energy is 
conserved  on both sides 
of the cut and 2) that the cut lines {\em exit} from the unshaded area and
{\em enter} into the shadowed one. Therefore the $\tau$-line accounts
for an arbitrary amount of energy entering (most likely) the unshadowed
area and exiting from the shadowed one, as indicated by the arrows in the
figure.}.
Namely the 4-momentum associated to a 
$\tau$-line is $\tau^\mu=(\tau,0,0,0)$. 
Following the prescriptions obtained 
in~\cite{veltman71}, the cut $\tau$-line 
introduces a factor $1/i\pi\times 1/\tau$
and a principal value integral is required. 
Then the latter diagrammatic equation writes 
to (apart from overall constant factors):
\begin{eqnarray}
\frac{1}{i\pi}\fint \frac{d\tau}{\tau} &&\left\{ 
\frac{\pi}{2}\frac{(q-\tau)^2-m^2}{(q-\tau)^2}\theta((q-\tau)^2-m^2) 
\right. \nonumber \\
&& + \left. \frac{\pi}
{2}\frac{(q+\tau)^2-m^2}{(q+\tau)^2} \theta((q+\tau)^2-m^2)\right\}. 
\end{eqnarray}
Let us make the shifts $q^0-\tau=-\tau^\prime\to \tau$ in the 
first integral and  $q^0+\tau=\tau^\prime\to\tau$ in the second to obtain
\begin{eqnarray}
\fint d\tau\left\{\frac{1}{q^0+\tau}\frac{\tau^2-{\bf q}^2-m^2}
{\tau^2-{\bf q}^2}\theta(\tau^2-{\bf q}^2-m^2)+\right.\nonumber\\ \left. 
\frac{1}{\tau-q^0}\frac{\tau^2-{\bf q}^2-m^2}{\tau^2-{\bf q}^2}\theta(\tau^2-{\bf q}^2-m^2)\right\}.
\label{shifft}
\end{eqnarray}
Apart from the $1/\tau$ factor, the cut diagrams are invariant 
and may be evaluated in any frame: we go to the frame ${\bf q}= {\bf 0}$, 
where the integral in (\ref{shifft}) reduces to
\begin{equation}
\fint\; d\tau\; \left\{\frac{2\tau}{\tau^2-q^{02}}
\frac{\tau^2-m^2}{\tau^2}\theta(\tau^2-m^2)\right\}.
\end{equation}
Setting $\tau^\prime=-\tau $ and $\tau^{\prime 2}=\lambda$ we find
\begin{equation}
\int_{m^2}^{\Lambda^2}\; d\lambda \; 
\frac{\lambda-m^2}{\lambda}\frac{1}{\lambda-q^2}.
\end{equation}
Solving this integral and using the fact that $\Lambda^2\gg q^2$ we find:
\begin{equation}
F-F^*=\ln \Lambda^2-\frac{q^2-m^2}{q^2}\ln (m^2-q^2)
-\frac{m^2}{q^2}\ln m^2 .
\label{imslf}
\end{equation}
With the proper constants taken into account we reconstruct the result obtained initially\footnote{Actually the two results differ by a constant
factor. This is a consequence of the fact that the chosen Green function
is divergent and the choice of two different UV regulators leads to
a constant shift in the lagrangian parameters.}.
Indeed from our initial calculation we get:
\begin{equation}
F-F^*=2\pi^2 i \left(\ln \Lambda^2-\frac{q^2-m^2}{q^2}
\ln (m^2-q^2)-\frac{m^2}{q^2}\ln m^2\right).
\end{equation}

\vskip0.5truecm

\begin{center}
\begin{figure}[hbt]
\SetScale{0.5}
\SetWidth{1.}
{\unitlength 0.5pt
\begin{picture}(0,0)(0,0)
\Line(180,0)(460,0)
\Line(220,0)(320,-100)
\Line(320,-100)(320,-140)
\Line(420,0)(320,-100)
\put(190,-30){$P_1$}
\put(450,-30){$P_2$}
\put(350,-125){$P_3$}
\end{picture} }
\vskip 80pt
\ccaption{}{\label{threept} Three point Green function in 
a scalar $\phi^3$ theory
}
\end{figure}
\end{center}

\subsection{The three point function}
\label{3points_sub}
We consider now a three point scalar function in $\phi^3$ theory. 
The Green function
depicted in Fig.~\ref{threept} is, up to constant factors, 
\[
G_3(P_1,P_2,P_3) \sim \int {\mathrm d^4 L} 
\frac{1}{[(L-P_1)^2-m^2] [(L-P_1-P_2)^2-m^2][L^2-m^2]},
\]
where $P_j\equiv(\Pi_j,\mathbf p_j)$
 denotes external momenta,
all momenta are assumed incoming,
  $m$ is the mass
of the scalar particle and $L\equiv(L_0,\mathbf l)$
is the loop four momentum and the standard
$L_0 \to L_0+i \epsilon$
prescription is understood. 

Using the rules explained above we can write:

\bea
G_3(P_1,P_2,P_3) & \sim & A_1+A_2+A_3+A_4\, , \\
A_1 & \sim & 
\int \frac {\mathrm d \tau} {\tau+i \epsilon}
\int \mathrm d \mathbf k_1  \mathrm d \mathbf k_2 
\delta (\Pi_1-\tau-E_1-E_2)\\ & &
\delta (  \mathbf p_1  -\mathbf k_1  
- \mathbf k_2 ) 
\frac{1}{E_1 E_2}
\frac{1}{[(K_1+P_3)^2-m^2]} \, ,\\
A_2 & \sim & 
\int \frac {\mathrm d \tau} {\tau+i \epsilon}
\int \mathrm d \mathbf k_1  \mathrm d \mathbf k_2
\delta (\Pi_2+\Pi_3-\tau-E_1-E_2)
\\ & &
\delta ( \mathbf p_2  +\mathbf p_3  
- \mathbf k_1  
- \mathbf k_2 ) 
\frac{1}{E_1 E_2}
\frac{1}{[(K_1-P_3)^2-m^2]} \, , \\
A_3 & \sim & 
\int \frac {\mathrm d \tau} {\tau+i \epsilon}
\int \mathrm d \mathbf k_1  \mathrm d \mathbf k_2
\delta (\Pi_1+\Pi_3-\tau-E_1-E_2)
\\ & &
\delta ( \mathbf p_1  +\mathbf p_3  
- \mathbf k_1  
- \mathbf k_2 ) 
\frac{1}{E_1 E_2}
\frac{1}{[(K_1-P_3)^2-m^2]} \, , \\
A_4 & \sim & 
\int \frac {\mathrm d \tau} {\tau+i \epsilon}
\int \mathrm d \mathbf k_1  \mathrm d \mathbf k_2
\delta (\Pi_2-\tau-E_1-E_2)
\\ & &
\delta ( \mathbf p_2  
- \mathbf k_1  
- \mathbf k_2 ) 
\frac{1}{E_1 E_2}
\frac{1}{[(K_1+P_3)^2-m^2]} \, ,
\eea
where $K_j\equiv(E_j,\mathbf k_j)=(\sqrt
{|\mathbf k_j|^2+m^2},\mathbf k_j)$

We can draw a $\tau$-line
between $P_1$ and $P_2$ vertices. We then cut
(put on mass-shell)
the corresponding internal lines and one of the other 
internal lines (this gives rise to two contributions
corresponding to $E_j$ both positive or negative).
We sum over all possible cuts and convolute with
the given weight as shown in Fig.~\ref{tauct}.
\begin{center}
\begin{figure}[hbt]
\SetScale{0.5}
\SetWidth{1.}
{\unitlength 0.5pt
\begin{picture}(0,0)(0,-100)
\Line(180,0)(460,0)
\Line(220,0)(320,-100)
\Line(320,-140)(320,-100)
\Line(420,0)(320,-100)
\put(190,-30){$P_1$}
\put(450,-30){$P_2$}
\put(350,-125){$P_3$}
\GlueArc(320,-80)(130,40,140){5}{8.57}
\begin{Large}
\put(540,-60){$=$}
\end{Large}
\ArrowLine(80,-220)(210,-220)
\ArrowLine(230,-220)(360,-220)
\ArrowLine(120,-220)(160,-260)
\ArrowLine(180,-280)(220,-320)
\Line(220,-320)(220,-360)
\Line(320,-220)(220,-320)
\Line(290,-150)(140,-300)
\Line(290,-150)(310,-170)
\Line(160,-320)(140,-300)

\Line(690,-150)(540,-300)
\Line(690,-150)(670,-130)
\Line(520,-280)(540,-300)

\put(90,-250){$P_1$}
\put(350,-250){$P_2$}
\put(250,-345){$P_3$}
\GlueArc(220,-300)(130,40,140){5}{8.57}
\begin{Large}
\put(200,-280){$A_1$}
\put(600,-280){$A_2$}
\put(200,-500){$A_3$}
\put(600,-500){$A_4$}
\put(400,-320){$+$}
\end{Large}
\ArrowLine(610,-220)(480,-220)
\ArrowLine(760,-220)(630,-220)
\ArrowLine(560,-260)(520,-220)
\ArrowLine(620,-320)(580,-280)
\Line(620,-360)(620,-320)
\Line(720,-220)(620,-320)
\put(490,-250){$P_1$}
\put(750,-250){$P_2$}
\put(650,-345){$P_3$}
\GlueArc(620,-300)(130,40,140){5}{8.57}
\end{picture} 
\begin {picture}(0,220)(0,120)
\ArrowLine(80,-220)(210,-220)
\ArrowLine(230,-220)(360,-220)
\Line(120,-220)(220,-320)
\Line(150,-150)(290,-290)
\Line(150,-150)(170,-130)
\Line(310,-270)(290,-290)

\Line(550,-150)(690,-290)
\Line(550,-150)(530,-170)
\Line(670,-310)(690,-290)

\Line(220,-320)(220,-360)
\ArrowLine(220,-320)(260,-280)
\ArrowLine(280,-260)(320,-220)
\put(90,-250){$P_1$}
\put(350,-250){$P_2$}
\put(250,-345){$P_3$}
\GlueArc(220,-300)(130,40,140){5}{8.57}
\begin{Large}
\put(400,-320){$+$}
\end{Large}
\ArrowLine(610,-220)(480,-220)
\ArrowLine(760,-220)(630,-220)
\Line(520,-220)(620,-320)
\Line(620,-320)(620,-360)
\ArrowLine(620,-320)(660,-280)
\ArrowLine(680,-260)(720,-220)
\put(490,-250){$P_1$}
\put(750,-250){$P_2$}
\put(650,-345){$P_3$}
\GlueArc(620,-300)(130,40,140){5}{8.57}
\end{picture} 
}
\vskip 270pt
\ccaption{}{\label{tauct} Three point Green function in 
a scalar $\phi^3$ theory. $\gluon$-line denotes the 
dispersive integral over $\mathrm d \tau$ as well as 
internal two body phase space. The arrows on internal
cut lines depict the energy flow in the cut diagram.
}
\end{figure}
\end{center}
Notice that we have obtained a dispersive formula:
each term is the convolution of two tree level Feynman
diagrams (with on-shell external particles and with a peculiar vertex
where energy conservation is guaranteed only by including the $\tau-$line contribution)
times two body phase space times $1/\tau \; d \tau$
integration very much like the dispersive integral of
the imaginary part of an amplitude.
This particular dispersive formulation
 is independent from the specific diagram.
 
We can go one step further. 
\begin{figure}[hbt]
\SetScale{0.5}
\SetWidth{1.}
{\unitlength 0.5pt
\begin{picture}(0,100)(0,0)
\Line(30,0)(70,0)
\put(50,-30){$P_1$}
\GlueArc(150,0)(80,145,180){5}{8.57}
\ArrowLine(70,0)(100,30)
\ArrowLine(70,0)(100,-30)
\begin{Large}
\put(120,-10){$*$}
\end{Large}
\begin{Huge}
\put(150,-10){$($}
\end{Huge}
\ArrowLine(190,30)(220,0)
\ArrowLine(190,-30)(220,0)
\Line(220,0)(260,0)
\Line(260,0)(290,30)
\Line(260,0)(290,-30)
\GlueArc(140,0)(80,0,35){5}{8.57}
\put(300,20){$P_2$}
\put(300,-40){$P_3$}
\begin{Large}
\put(330,-10){$+$}
\end{Large}
\ArrowLine(380,30)(420,30)
\ArrowLine(380,-30)(420,-30)
\Line(420,-30)(420,30)
\Line(420,-30)(460,-30)
\Line(420,30)(460,30)
\GlueArc(390,0)(43,45,110){5}{8.57}
\put(470,20){$P_2$}
\put(470,-40){$P_3$}
\begin{Large}
\put(500,-10){$+$}
\end{Large}
\end{picture} 
\begin{picture}(-170,0)(-170,0)
\ArrowLine(380,30)(420,30)
\ArrowLine(380,-30)(420,-30)
\Line(420,-30)(420,30)
\Line(420,-30)(460,-30)
\Line(420,30)(460,30)
\GlueArc(390,0)(43,45,110){5}{8.57}
\put(470,20){$P_3$}
\put(470,-40){$P_2$}
\begin{Huge}
\put(500,-10){$)$}
\end{Huge}
\end{picture} 
}
\vskip 40pt
\ccaption{}{\label{amplitude}
Convolution of amplitudes rather than individual Feynman graphs.
It is manifest that, in addition to vertex contributions, we pick
up also self energies contributions.
}
\end{figure}
Let us focus
on the contribution $A_1$: it is the convolution of the 
process 
\[
\phi_1  \to \phi_1^{(cut)} \phi_2^{(cut)}
\]
and the $t$-channel graph for the process 
\beq
 \phi_1^{(cut)} \phi_2^{(cut)} \to \phi_2\phi_3\, ,
\label{prc1a}
\eeq
where $\phi_j$ denotes external particles with momentum
$P_j$ and  $\phi_j^{(cut)}$ denotes internal (cut)
particles.
If we extend $A_1$ to include the contribution of the
whole amplitude ($s$, $t$, and $u$ channel contributions,
see Fig.~\ref{amplitude}) of process (\ref{prc1a}),
we also obtain the contribution 
of the $\phi_1$ external self-energies as well as part of the 
triangles for the process 
\beq
\phi_1 \to \phi_2 \phi_3\, .
\label{proc}
\eeq
Applying the same procedure to all the $A_j$ graphs, we obtain the one-loop
correction to the amplitude for the process (\ref{proc}) as given by the 
formula 
\bea
\phi_1 \to \phi_2 \phi_3
& \equiv &
\sum_{\buildrel{j_1,j_2,j_3=1,3}\over{j_1\ne j_2, j_1\ne j_3,
j_2 \ne j_3}}
(\phi_{j_1}
 \phi_{j_2} \to \phi_1^{(cut)} \phi_2^{(cut)} ) *  
(\phi_1^{(cut)} \phi_2^{(cut)} \to \phi_{j_3}) \\
& & +
(\phi_{j_1} \to \phi_1^{(cut)} \phi_2^{(cut)}) *  
(\phi_1^{(cut)} \phi_2^{(cut)} \to  \phi_{j_2} \phi_{j_3} )\, ,
\eea
where with the symbol $*$ we denote the convolution with
the two body phase space and the dispersive integral 
over $\mathrm d \tau$. Actually, to define
unambiguously  the above expression one needs a prescription
to pick up the two vertices where energy conservation 
is ensured by the $\tau$-line contribution, and one needs to check that 
each diagram is accounted for with the correct combinatorial factors. 
This will be discussed later on. 

{\em We have thus achieved our first goal:} with the algorithm 
sketched above we can write any one loop amplitude as
a convolution in 4-dimensions of tree level amplitudes. More precisely, 
given a set of external particles, we divide it into
two non empty subsets ${\mathcal G}_{\alpha_j}$ $j=1,2$. We
compute the convolution of
\beq
({\mathcal G}_{\alpha_1} \to \chi_1 \chi_2) *
(\chi_1 \chi_2 \to {\mathcal G}_{\alpha_2}  )\, ,
\label{algorithm}
\eeq
summing over all possible intermediate particles $\chi_j$
(here ``possible'' means compatible with the symmetries of 
the theory). Recall that tree-level amplitudes are computed
for on-shell particles but an arbitrary amount of energy 
can outflow/sink from/in one vertex. Therefore
one of the two amplitudes 
in (\ref{algorithm}) can be non zero also
for all particles incoming/outgoing. Finally we sum
over all possible ${\mathcal G}_{\alpha_1}$, 
${\mathcal G}_{\alpha_2}$ partitions.

\section{The general algorithm}
In this section we outline the general algorithm we use to draw the 
$\tau$-lines and to select all the cut diagrams with
the proper weights and combinatorial factors.

For definiteness we consider the amplitude
\ben
\phi(P_1) \phi(P_2) \to \phi(P_3) \phi(P_4)
\label{two_to_two}
\een
in $\phi^3$ theory.
 $P_j$ denote, as before,
the momenta of the external particles.

\bit
\item We divide the set of external particles into
two non empty subsets 
${\mathcal G}_{\alpha_j}$ $j=1,2$. There are 14 
$\left(=\sum_{i=1}^3\pmatrix{4\cr i}\right)$ such partitions.

\item
We compute the convolution of
\ben
[{\mathcal G}_{\alpha_1} \to \phi(K_1) \phi(K_2) ]*[
\phi(K_1) \phi(K_2) \to {\mathcal G}_{\alpha_2}  ]
\label{convolution}
\een
summing over  all possible partitions. We have denoted with 
$K_j$ the momenta of the cut lines.

\item Require that the $\tau$ line is
emitted/absorbed at the opposite ends of the cut-line
carrying $K_1$ momentum.

\item Select an external momentum (arbitrarily). For
definiteness we choose $P_1$. We will call it reference momentum from here on.

\item If the reference momentum belongs to
 ${\mathcal G}_{\alpha_2}$ the amplitude for \\
${\mathcal M}_2 = \phi(K_1) \phi(K_2) \to {\mathcal G}_{\alpha_2}$ needs to 
be modified as follows: 

\bit

\item If a diagram has a propagator depending on a momentum 
$Q = \sum_j Q_j$, where $Q_j \neq K_1 $ and at least $P_1$ and $K_2$ 
enter the sum over $j$, this diagram
must be vetoed.

\item If a diagram has a propagator containing
$K_1$ and at least an external $P_j$ but not $P_1$, this diagram
must be vetoed.

\item The diagram of ${\mathcal M}_2$ with $K_1$ and $K_2$ attached 
to the same vertex should be vetoed if ${\mathcal G}_{\alpha_1}$ 
is made up of a single external particle. This prescription allows us 
to avoid to compute external particle self energies. 
This contributions will be computed analytically and added in a second step.

\eit

\item The above prescriptions will apply also to
${\mathcal G}_{\alpha_1}
\to \phi(K_1) \phi(K_2) $ if $P_1$ belongs
 to ${\mathcal G}_{\alpha_1}$. 
\eit
Let us examine one of such partitions:
\[
{\mathcal G}_{\alpha_1}=\{ \phi(P_4) \}; \quad
{\mathcal G}_{\alpha_2}=\{ \phi(P_1),\phi(P_2),\phi(P_3) \} 
\]The calculation of the convolution ${\mathcal M}_{\alpha_1} 
* {\mathcal M}_{\alpha_2}$ for such partition 
is sketched in Fig.~\ref{amplitude2}. 
Vetoed Feynman diagrams are marked in the figure. 
Summing up over all fourteen partitions one can check 
by inspection that all contributions to NLO correction
are correctly taken into account and that any relevant cut
occurs the correct number of times.

After providing the recipe, let us try to motivate it.
We shall discuss one of the box diagrams 
contributing to the amplitude in Eqn.~(\ref{two_to_two}).
As shown in fig.~\ref{box} there are six possible different ways
to draw the $\tau$-line, four corresponding to the $\tau$-line
emitted and absorbed by two nearby vertices and two corresponding
 to the $\tau$-line
emitted and absorbed by two opposite vertices.

\begin{center}
\begin{figure}[hbt]
\SetScale{0.4}
\SetWidth{1.}
{\unitlength 0.4pt
\begin{picture}(0,0)(100,50)
\Line(180,0)(460,0)
\Line(220,0)(220,-200)
\Line(420,0)(420,-200)
\Line(180,-200)(460,-200)
\put(190,-30){$P_1$}
\put(450,-30){$P_2$}
\put(190,-230){$P_4$}
\put(450,-230){$P_3$}
\GlueArc(320,-80)(130,40,140){5}{8.57}
\DashLine(320,80)(320,-250){5}
\put(290,-150){$\buildrel { {\mathcal C}_1 } \over \leftarrow$}
\put(330,-150){$\buildrel { {\mathcal C}_2 } \over \rightarrow$}
\DashLine(140,-80)(420,60){5}
\put(140,-65){$\buildrel { {\mathcal C}_3 } \over \leftarrow$}
\put(160,-100){$\buildrel { {\mathcal C}_4 } \over \rightarrow$}
\DashLine(520,-80)(240,60){5}
\put(500,-65){$\buildrel { {\mathcal C}_5 } \over \leftarrow$}
\put(485,-100){$\buildrel { {\mathcal C}_6 } \over \rightarrow$}
\end{picture} 
\begin{picture}(0,0)(-400,50)
\Line(180,0)(460,0)
\Line(220,0)(220,-200)
\Line(420,0)(420,-200)
\Line(180,-200)(460,-200)
\put(190,-30){$P_1$}
\put(450,-30){$P_2$}
\put(190,-230){$P_4$}
\put(450,-230){$P_3$}
\DashLine(320,80)(320,-250){5}
\put(290,-150){$\buildrel { {\mathcal C}_1 } \over \leftarrow$}
\put(330,-150){$\buildrel { {\mathcal C}_2 } \over \rightarrow$}
\DashLine(140,-80)(420,60){5}
\put(140,-65){$\buildrel { {\mathcal C}_3 } \over \leftarrow$}
\put(160,-100){$\buildrel { {\mathcal C}_4 } \over \rightarrow$}
\DashLine(240,-230)(520,-90){5}
\put(240,-215){$\buildrel { {\mathcal C}_5 } \over \leftarrow$}
\put(260,-250){$\buildrel { {\mathcal C}_6 } \over \rightarrow$}
\DashLine(170,-100)(470,-100){5}

\put(240,-80){$ { {\mathcal C}_7 }  \uparrow$}
\put(240,-120){$ { {\mathcal C}_8 }  \downarrow$}

\Gluon(220,0)(420,-200){6}{11.1}
\end{picture} 

}
\vskip 150pt
\ccaption{}{\label{box}
Four point function. There are six possible ways to draw the $\tau$-line: four
as shown in the left-hand side of the figure (the other three possibilities
are similar but with the $\tau$-line among $P_2$ and $P_3$ or  
$P_3$ and $P_4$ or  $P_4$ and $P_1$) and two as shown in the 
right-hand side of the figure (the other one is
similar but with the $\tau$-line among $P_2$ and $P_4$).
The left hand choice gives the sum of six contributions
corresponding to having the propagator connecting $P_1$ and $P_2$
 cut together with anyone of the remaining propagators.
The right hand choice gives eight contributions 
corresponding to having a cut propagator between $P_1$ and $P_2$ or 
between $P_2$ and $P_3$,  together with 
the propagator between $P_3$ and $P_4$ or 
between $P_4$ and $P_1$. Recall that each ``cut'' actually corresponds 
to two contributions depending on the energy flow. 
}
\end{figure}
\end{center}
Let us focus on the case of the  $\tau$-line between two near-by vertices:
\footnote{The other choice is possible as well. It requires a somewhat
more involved algorithm. We have not studied it insofar since 
at present we do not have
a good reason to believe it leads to easier integrals.} the diagrams 
will be written as the sum over six possible cuts.
If we denote by $\pi_{ij}$ the propagator connecting $P_i$ and $P_j$
and by $(\pi_{jk},\pi_{lm})$ the contribution corresponding
to  $\pi_{jk}$ and $\pi_{lm}$ cut propagators,
the left-hand diagram of fig.~\ref{box} will result from the sum of the 
$(\pi_{12},\pi_{41})$, $(\pi_{12},\pi_{34})$ and $(\pi_{12},\pi_{23})$
cuts, each counting as two distinct cuts dependig on the energy flow.
 Each of these 
contributions will be included into a different term of the convolution
of Eqn.~(\ref{convolution}), namely it will arise with a different choice
of $\alpha_j$ (notice that on the opposite sides of the cuts 
in Fig.~\ref{box} there are different sets of external particles). Therefore we
need a prescription to ensure that all the cuts enter
into the final sum with the same combinatorial factor and that we do not 
have contributions from the other possible ways of drawing the $\tau$-line.

The simplest prescription is to require that the $\tau$-line
is emitted and absorbed together with particle $\phi(K_1)$, namely
it will enter/exit always from the vertices to which $\phi(K_1)$ is
attached. In this way the $\tau$-line is always emitted and reabsorbed
at the opposite side of {\em one} propagator and we single out the topology
we are after (left panel of Fig.~\ref{box}).

If we allow $\phi(K_1)$ to connect to all possible external particles, 
the diagram of Fig.~\ref{box} (left panel) would be reconstructed four 
times, since the $\tau$-line would connect all possible near-by vertices. 
With the same reasoning it's
easy to conclude that a n-point diagram will be recovered n times.

To adjust the combinatorial factors, we introduce the reference
momentum $P_1$: we require that $\phi(K_1)$ is attached to the same
vertex with $\phi(P_1)$ or with a tree of external particles
containing $\phi(P_1)$ as shown in fig.~\ref{refmom}. Notice that in this 
way we obtain each diagram twice (the $\phi(K_1)$ vertex has {\em two} 
near-by vertices) with the exception of self-energy contributions which
are generated only once. These are indeed the correct combinatorial 
factors.

\begin{center}
\begin{figure}[hbt]
\SetScale{0.5}
\SetWidth{1.}
{\unitlength 0.5pt
\begin{picture}(0,0)(0,50)
\Line(180,0)(320,0)
\Line(220,0)(220,-60)
\Line(270,0)(270,-60)
\GlueArc(250,-30)(45,140,180){5}{4.57}
\put(160,-10){$P_1$}
\put(215,-90){$K_1$}
\put(265,-90){$P_2$}
\put(325,-10){$K_2$}
\CCirc(310,0){5}{Black}{Black}
\CCirc(220,-50){5}{Black}{Black}
\end{picture} 
\begin{picture}(0,0)(-300,50)
\Line(130,0)(320,0)
\Line(180,0)(180,-60)
\Line(220,0)(220,-60)
\Line(270,0)(270,-60)
\GlueArc(250,-30)(45,140,180){5}{4.57}
\put(110,-10){$P_1$}
\put(215,-90){$K_1$}
\put(165,-90){$P_3$}
\put(265,-90){$P_2$}
\put(325,-10){$K_2$}
\CCirc(310,0){5}{Black}{Black}
\CCirc(220,-50){5}{Black}{Black}
\end{picture} 
}
\vskip 150pt
\ccaption{}{\label{refmom}
Allowed particles tree into the amplitudes entering into the convolution
of Eqn.~(\ref{convolution}). Cut propagators (denoted with a small circle)
carry momentum $K_j$
and external particles carry momentum $P_j$. The $\tau$-line is always
emitted/absorbed from a vertex connected
 either directly to $P_1$ (left-hand side) or to 
a tree made up of external particles only ($K_2$ not allowed) and
containing $P_1$. Equivalently, if the tree is seen from the opposite side,
the $\tau$-line is always 
emitted/absorbed from a vertex connected to a tree  
made up of external particles plus $K_2$ and not containing $P_1$.
The left-hand side diagram, inside the convolution, will contribute to
the evaluation of boxes and triangles, whereas the right-hand side
to triangles only.
}
\end{figure}
\end{center}
All the prescriptions given above to veto certain contributions to
the amplitudes aim at implementing the above reasoning. 
They amount to require that any particle tree starting from $\phi(P_1)$,
to be accepted,
should absorbe $\phi(K_1)$ {\em before}  $\phi(K_2)$, as shown in 
Figs.~(\ref{refmom}) and (\ref{veto}).
They look somewhat
involved since, due to four-momentum conservation, 
the momentum flowing into a given propagator can 
be computed via two possible combinations of particles 
and we do not know a priori which will be chosen from our tool 
for the automatic evaluation of tree-level amplitudes. Therefore 
we cannot simply check if $K_1$ and $P_1$ enter into the "same" 
 (in the extended sense described above) vertex, but we need also 
to check (and accept) also trees made up of external particles 
$\phi(P_j)$ $(j \neq 1)$ and/or $\phi(K_2)$
\begin{center}
\begin{figure}[hbt]
\SetScale{0.5}
\SetWidth{1.}
{\unitlength 0.5pt
\begin{picture}(0,0)(0,50)
\Line(180,0)(320,0)
\Line(220,0)(220,-60)
\Line(270,0)(270,-60)
\GlueArc(250,-30)(45,140,180){5}{4.57}
\put(160,-10){$P_2$}
\put(215,-90){$K_1$}
\put(265,-90){$P_1$}
\put(325,-10){$K_2$}
\CCirc(310,0){5}{Black}{Black}
\CCirc(220,-50){5}{Black}{Black}
\end{picture} 
\begin{picture}(0,0)(-250,50)
\Line(180,0)(370,0)
\Line(220,0)(220,-60)
\Line(270,0)(270,-60)
\Line(320,0)(320,-60)
\GlueArc(250,-30)(45,140,180){5}{4.57}
\put(160,-10){$P_2$}
\put(315,-90){$P_1$}
\put(215,-90){$K_1$}
\put(265,-90){$P_3$}
\put(375,-10){$K_2$}
\CCirc(360,0){5}{Black}{Black}
\CCirc(220,-50){5}{Black}{Black}
\end{picture} 

}
\vskip 100pt
\ccaption{}{\label{veto}
Vetoed particles tree into the amplitudes entering into the convolution
of Eqn.~(\ref{convolution}). Cut propagators (denoted with a small circle)
carry momentum $K_j$
and external particles carry momentum $P_j$. The $K_2$ particle is always
emitted/absorbed from a vertex connected
 either directly to $P_1$ (left-hand side) or to 
a tree made up of external particle only ($K_1$ not allowed) and
containing $P_1$. Equivalently, if the tree is seen from the opposite side, 
the $K_1$ particle is always
emitted/absorbed from a vertex connected to a tree,  
made up of external particles, which does not contain 
neither $K_2$ nor $P_1$.
}
\end{figure}
\end{center}
Coming back to the four point function we sketch in fig.~\ref{box1}
the way it arises from the convolution in Eqn.~(\ref{convolution}).
For simplicity we limit to a single box contributing to the
amplitude and denote with dots all the remaining contributions.
As shown  in fig.~\ref{box1} the box is evaluated twice corresponding to
two different ways of drawing the $\tau$-line.
\begin{center}
\begin{figure}[hbt]
\SetScale{0.5}
\SetWidth{1.}
{\unitlength 0.5pt
\begin{picture}(0,0)(100,50)
\Line(220,0)(370,-150)
\Line(220,-150)(370,0)
\GCirc(295,-75){40}{0.8}
\put(180,-30){$P_1$}
\put(380,-30){$P_2$}
\put(180,-180){$P_4$}
\put(380,-180){$P_3$}
\begin{Huge}
\put (140,-90){$ 4 $}
\end {Huge}
\put (410,-90){$= \sum_{j}
[{\mathcal G}_{\alpha_j} \to \phi(K_1) \phi(K_2) ]*[
\phi(K_1) \phi(K_2) \to {\mathcal G}_{\bar \alpha_j}  ]$}
\end{picture}
\begin{picture}(0,0)(100,260)
\Line(180,0)(390,0)
\Line(210,0)(210,-150)
\Line(360,0)(360,-150)
\Line(180,-150)(390,-150)
\put(180,-30){$P_1$}
\put(360,-30){$P_2$}
\put(180,-180){$P_4$}
\put(360,-180){$P_3$}
\begin{Huge}
\put (120,-90){$4$ }
\put (410,-90){$+ \dots = $ }
\end {Huge}
\end{picture} 
\begin{picture}(0,0)(100,470)
\Line(180,0)(390,0)
\Line(210,0)(210,-150)
\Line(360,0)(360,-150)
\Line(180,-150)(390,-150)
\put(180,-30){$P_1$}
\put(360,-30){$P_2$}
\put(180,-180){$P_4$}
\put(360,-180){$P_3$}
\DashLine(285,40)(285,-190){5}
\put(255,-85){$\buildrel { {\mathcal C}_3 } \over \leftarrow$}
\put(295,-85){$\buildrel { {\mathcal C}_4 } \over \rightarrow$}
\DashLine(159,-60)(369,40){5}
\put(145,-55){$\buildrel { {\mathcal C}_1 } \over \leftarrow$}
\put(165,-90){$\buildrel { {\mathcal C}_2 } \over \rightarrow$}
\DashLine(411,-60)(201,40){5}
\put(425,-55){$\buildrel { {\mathcal C}_5 } \over \rightarrow$}
\put(395,-90){$\buildrel { {\mathcal C}_6 } \over \leftarrow$}
\GlueArc(285,-70)(100,40,140){5}{8.57}
\put(265,-105){${\mathcal D}_1$}

\begin{Huge}
\put (420,-90){$+$}
\end {Huge}

\end{picture} 
\begin{picture}(0,0)(-250,470)
\Line(180,0)(390,0)
\Line(210,0)(210,-150)
\Line(360,0)(360,-150)
\Line(180,-150)(390,-150)
\put(180,-30){$P_1$}
\put(360,-30){$P_2$}
\put(180,-180){$P_4$}
\put(360,-180){$P_3$}

\DashLine(170,-75)(390,-75){5}
\put(310,-105){$ { {\mathcal C}_{10} }  \downarrow$}
\put(310,-65){$ { {\mathcal C}_{9} }  \uparrow$}
\DashLine(170,-135)(280,30){5}
\put(250,5){$\buildrel { {\mathcal C}_7 } \over \leftarrow$}
\put(290,5){$\buildrel { {\mathcal C}_8 } \over \rightarrow$}
\DashLine(170,-15)(280,-180){5}
\put(250,-190){$\buildrel { {\mathcal C}_{11} } \over \leftarrow$}
\put(290,-190){$\buildrel { {\mathcal C}_{12} } \over \rightarrow$}

\GlueArc(285,-75)(100,130,230){5}{8.57}
\put(265,-105){${\mathcal D}_2$}

\begin{Huge}
\put (410,-90){$+ \dots $ }
\end {Huge}

\end{picture} 
\begin{picture}(0,0)(0,680)
\put(0,0){
}
\end{picture} 
}
\vskip 335pt
\ccaption{}{\label{box1}Cuts contributing to the four point function. Only 
those relevant to a specific box are shown}

\end{figure}
\end{center}
 The various cuts,
denoted with ${\mathcal C}_j$ in fig.~\ref{box1}, arise as follows in 
Eqn.~(\ref{convolution})
\bea
[{\mathcal G}_{1} \to \phi(K_1) \phi(K_2) ] 
*
 [
\phi(K_1) \phi(K_2) \to {\mathcal G}_{2,3,4}  ] &= &
{\mathcal C}_2 +{\mathcal C}_8 + \dots 
\\
\left [{\mathcal G}_{2} \to \phi(K_1) \phi(K_2) \right ] * [
\phi(K_1) \phi(K_2) \to {\mathcal G}_{1,3,4}  ] &= &
{\mathcal C}_6  + \dots 
\\
\left [{\mathcal G}_{3} \to \phi(K_1) \phi(K_2) \right ] * [
\phi(K_1) \phi(K_2) \to {\mathcal G}_{1,2,4}  ] &= &
   \dots 
\\
\left [{\mathcal G}_{4} \to \phi(K_1) \phi(K_2) \right ] * [
\phi(K_1) \phi(K_2) \to {\mathcal G}_{1,2,3}  ] &= &
 {\mathcal C}_{12}+  \dots 
\\
\left [{\mathcal G}_{1,2} \to \phi(K_1) \phi(K_2) \right ] * [
\phi(K_1) \phi(K_2) \to {\mathcal G}_{3,4}  ] &= &
 {\mathcal C}_{10}+  \dots 
\\
\left [{\mathcal G}_{1,3} \to \phi(K_1) \phi(K_2) \right ] * [
\phi(K_1) \phi(K_2) \to {\mathcal G}_{2,4}  ] &= &
  \dots 
\\
\left [{\mathcal G}_{1,4} \to \phi(K_1) \phi(K_2) \right ] * [
\phi(K_1) \phi(K_2) \to {\mathcal G}_{2,3}  ] &= &
 {\mathcal C}_{4}+  \dots 
\\
\left [{\mathcal G}_{2,3} \to \phi(K_1) \phi(K_2) \right ] * [
\phi(K_1) \phi(K_2) \to {\mathcal G}_{1,4}  ] &= &
 {\mathcal C}_{3}+  \dots 
\\
\left [{\mathcal G}_{2,4} \to \phi(K_1) \phi(K_2) \right ] * [
\phi(K_1) \phi(K_2) \to {\mathcal G}_{1,3}  ] &= &
 \dots 
\\
\left [{\mathcal G}_{3,4} \to \phi(K_1) \phi(K_2) \right ] * [
\phi(K_1) \phi(K_2) \to {\mathcal G}_{1,2}  ] &= &
 {\mathcal C}_{9}+  \dots 
\\
\left [{\mathcal G}_{1,2,3} \to \phi(K_1) \phi(K_2) \right ] * [
\phi(K_1) \phi(K_2) \to {\mathcal G}_{4}  ] &= &
 {\mathcal C}_{11}+  \dots 
\\
\left [{\mathcal G}_{1,2,4} \to \phi(K_1) \phi(K_2) \right ] * [
\phi(K_1) \phi(K_2) \to {\mathcal G}_{3}  ] &= &
  \dots 
\\
\left [{\mathcal G}_{1,3,4} \to \phi(K_1) \phi(K_2) \right ] * [
\phi(K_1) \phi(K_2) \to {\mathcal G}_{2}  ] &= &
 {\mathcal C}_{5}+  \dots 
\\
\left [{\mathcal G}_{2,3,4} \to \phi(K_1) \phi(K_2) \right ] * [
\phi(K_1) \phi(K_2) \to {\mathcal G}_{1}  ] &= &
 {\mathcal C}_{1}+{\mathcal C}_{7} \dots 
\eea

The above algorithm applies also to more complex cases with
one important addition:
\bit

\item If cut particles are not identical, one needs
to consider both the contribution with
$K_1 \leftrightarrow K_2$. Namely the contribution of
$\tau$-line
absorbed/emitted from $K_1$ with the above vetoing
prescriptions
plus the contribution of
$\tau$-line
absorbed/emitted from $K_2$ with the above vetoing
prescriptions but with
$K_1 \leftrightarrow K_2$.
The sum of this two contributions leads to twice
the results and again one can check by direct inspection
that the combinatorics of cut diagrams is correct.
\eit

The entire procedure is numerical: we perform a numerical 
integration {\em simultaneously} over the {\em phase space variables
of external particles}, over $d \tau$ and
{\em phase space variables of intermediate particles}. Each
computation
is performed in four space-time dimensions. If needed
the integral in  $d \tau$ is truncated at
$\tau<\tau_{\mathrm max}$ (where $\tau_{\rm max}$ is an 
ultraviolet cutoff) and proper counterterms are included 
to ensure that the integral is properly regularized. 
In  practice one computes a subtracted Green function 
and then modifies the tree-level Lagrangian with 
appropriate counterterms in  such a way to reproduce 
the correct results for divergent Green functions.

\begin{center}
\begin{figure}[hbt]
\SetScale{0.5}
\SetWidth{1.}
{\unitlength 0.5pt
\begin{picture}(0,100)(0,0)
\Line(30,0)(70,0)
\put(0,0){$P_4$}
\GlueArc(150,0)(80,145,180){5}{8.57}
\ArrowLine(70,0)(100,30)
\ArrowLine(70,0)(100,-30)
\begin{Huge}
\put(120,-10){$*$}
\end{Huge}
\end{picture}
\begin{picture}(0,0)(100,100)
\ArrowLine(190,30)(220,0)
\ArrowLine(190,-30)(220,0)
\Line(220,0)(300,0)
\Line(300,0)(340,30)
\Line(300,0)(340,-30)
\Line(260,0)(300,30)
\GlueArc(140,0)(80,0,35){5}{8.57}
\put(340,40){$P_2$}
\put(340,-60){$P_3$}
\put(300,40){$P_1$}
\begin{Large}
\put(360,-10){$+$}
\end{Large}
\begin{Huge}
\put(140,-10){$($}
\end{Huge}
\end{picture}
\begin{picture}(0,0)(-100,100)
\ArrowLine(190,30)(220,0)
\ArrowLine(190,-30)(220,0)
\Line(220,0)(300,0)
\Line(300,0)(340,30)
\Line(300,0)(340,-30)
\Line(260,0)(300,30)
\GlueArc(140,0)(80,0,35){5}{8.57}
\put(340,40){$P_3$}
\put(340,-60){$P_1$}
\put(300,40){$P_2$}
\begin{Large}
\put(360,-10){$+$}
\end{Large}
\end{picture}
\begin{picture}(0,0)(-300,100)
\ArrowLine(190,30)(220,0)
\ArrowLine(190,-30)(220,0)
\Line(220,0)(300,0)
\Line(300,0)(340,30)
\Line(300,0)(340,-30)
\Line(260,0)(300,30)
\GlueArc(140,0)(80,0,35){5}{8.57}
\put(340,40){$P_1$}
\put(340,-60){$P_2$}
\put(300,40){$P_3$}
\begin{Large}
\put(360,-10){$+$}
\end{Large}
\end{picture}
\begin{picture}(0,0)(120,250)
\ArrowLine(190,30)(220,0)
\Line(190,-30)(220,0)
\Line(220,0)(300,0)
\Line(300,0)(340,30)
\Line(300,0)(340,-30)
\ArrowLine(300,30)(260,0)
\GlueArc(140,0)(80,0,35){5}{8.57}
\put(340,40){$P_2$}
\put(340,-60){$P_3$}
\put(180,-60){$P_1$}
\begin{Large}
\put(360,-10){$+$}
\end{Large}
\end{picture}
\begin{picture}(0,0)(-80,250)
\ArrowLine(190,30)(220,0)
\Line(190,-30)(220,0)
\Line(220,0)(300,0)
\Line(300,0)(340,30)
\Line(300,0)(340,-30)
\ArrowLine(300,30)(260,0)
\GlueArc(140,0)(80,0,35){5}{8.57}
\put(340,40){$P_3$}
\put(340,-60){$P_1$}
\put(180,-60){$P_2$}
\put(15,120){vetoed}
\put(230,120){vetoed}
\put(430,120){vetoed}
\put(30,-30){$a_1$}
\put(230,-30){vetoed}
\put(430,-30){vetoed}
\put(30,-180){$a_2$}
\put(230,-180){vetoed}
\put(430,-180){vetoed}
\put(30,-330){$a_3$}
\put(230,-330){vetoed}
\put(430,-330){vetoed}
\put(15,-480){vetoed}
\put(245,-480){$a_4$}
\put(445,-480){$a_5$}
\begin{Large}
\put(360,-10){$+$}
\end{Large}
\end{picture}
\begin{picture}(0,0)(-280,250)
\ArrowLine(190,30)(220,0)
\Line(190,-30)(220,0)
\Line(220,0)(300,0)
\Line(300,0)(340,30)
\Line(300,0)(340,-30)
\ArrowLine(300,30)(260,0)
\GlueArc(140,0)(80,0,35){5}{8.57}
\put(340,40){$P_1$}
\put(340,-60){$P_2$}
\put(180,-60){$P_3$}
\begin{Large}
\put(360,-10){$+$}
\end{Large}
\end{picture}
\begin{picture}(0,0)(140,400)
\ArrowLine(190,30)(220,0)
\Line(190,-30)(220,0)
\Line(220,0)(300,0)
\Line(300,0)(340,-30)
\Line(300,30)(260,0)
\ArrowLine(340,30)(300,0)
\GlueArc(140,0)(80,0,35){5}{8.57}
\put(300,40){$P_2$}
\put(340,-60){$P_3$}
\put(180,-60){$P_1$}
\begin{Large}
\put(360,-10){$+$}
\end{Large}
\end{picture}
\begin{picture}(0,0)(-60,400)
\ArrowLine(190,30)(220,0)
\Line(190,-30)(220,0)
\Line(220,0)(300,0)
\ArrowLine(340,30)(300,0)
\Line(300,0)(340,-30)
\Line(300,30)(260,0)
\GlueArc(140,0)(80,0,35){5}{8.57}
\put(300,40){$P_3$}
\put(340,-60){$P_1$}
\put(180,-60){$P_2$}
\begin{Large}
\put(360,-10){$+$}
\end{Large}
\end{picture}
\begin{picture}(0,0)(-260,400)
\ArrowLine(190,30)(220,0)
\Line(190,-30)(220,0)
\Line(220,0)(300,0)
\ArrowLine(340,30)(300,0)
\Line(300,0)(340,-30)
\Line(300,30)(260,0)
\GlueArc(140,0)(80,0,35){5}{8.57}
\put(300,40){$P_1$}
\put(340,-60){$P_2$}
\put(180,-60){$P_3$}
\begin{Large}
\put(360,-10){$+$}
\end{Large}
\begin{Large}
\put(360,-10){$+$}
\end{Large}
\end{picture}
\begin{picture}(0,0)(160,550)
\ArrowLine(190,30)(220,0)
\Line(190,-30)(220,0)
\Line(220,0)(300,0)
\Line(300,0)(340,-30)
\Line(300,30)(260,0)
\ArrowLine(340,30)(300,0)
\GlueArc(140,0)(80,0,35){5}{8.57}
\put(300,40){$P_3$}
\put(340,-60){$P_2$}
\put(180,-60){$P_1$}
\begin{Large}
\put(360,-10){$+$}
\end{Large}
\end{picture}
\begin{picture}(0,0)(-40,550)
\ArrowLine(190,30)(220,0)
\Line(190,-30)(220,0)
\Line(220,0)(300,0)
\ArrowLine(340,30)(300,0)
\Line(300,0)(340,-30)
\Line(300,30)(260,0)
\GlueArc(140,0)(80,0,35){5}{8.57}
\put(300,40){$P_2$}
\put(340,-60){$P_1$}
\put(180,-60){$P_3$}
\begin{Large}
\put(360,-10){$+$}
\end{Large}
\end{picture}
\begin{picture}(0,0)(-240,550)
\ArrowLine(190,30)(220,0)
\Line(190,-30)(220,0)
\Line(220,0)(300,0)
\ArrowLine(340,30)(300,0)
\Line(300,0)(340,-30)
\Line(300,30)(260,0)
\GlueArc(140,0)(80,0,35){5}{8.57}
\put(300,40){$P_1$}
\put(340,-60){$P_3$}
\put(180,-60){$P_2$}
\begin{Large}
\put(360,-10){$+$}
\end{Large}
\end{picture}
\begin{picture}(0,0)(180,700)
\ArrowLine(190,30)(220,0)
\Line(190,-30)(220,0)
\Line(220,0)(300,0)
\Line(300,0)(340,-30)
\ArrowLine(300,30)(260,0)
\Line(340,30)(300,0)
\GlueArc(340,0)(80,145,180){5}{8.57}
\put(340,40){$P_2$}
\put(340,-60){$P_3$}
\put(180,-60){$P_1$}
\begin{Large}
\put(360,-10){$+$}
\end{Large}
\end{picture}
\begin{picture}(0,0)(-20,700)
\ArrowLine(190,30)(220,0)
\Line(190,-30)(220,0)
\Line(220,0)(300,0)
\Line(340,30)(300,0)
\Line(300,0)(340,-30)
\ArrowLine(300,30)(260,0)
\GlueArc(340,0)(80,145,180){5}{8.57}
\put(340,40){$P_3$}
\put(340,-60){$P_1$}
\put(180,-60){$P_2$}
\begin{Large}
\put(360,-10){$+$}
\end{Large}
\end{picture}
\begin{picture}(0,0)(-220,700)
\ArrowLine(190,30)(220,0)
\Line(190,-30)(220,0)
\Line(220,0)(300,0)
\Line(340,30)(300,0)
\Line(300,0)(340,-30)
\ArrowLine(300,30)(260,0)
\GlueArc(340,0)(80,145,180){5}{8.57}
\put(340,40){$P_1$}
\put(340,-60){$P_2$}
\put(180,-60){$P_3$}
\begin{Huge}
\put(370,-10){$)$}
\end{Huge}
\end{picture}
}
\vskip 400pt
\ccaption{}{\label{amplitude2}
Convolution of amplitudes rather than individual Feynman graphs. 
Cut lines carry an arrow. $\gluon$ indicates a $\tau$-line. 
The graphs contained in the picture correspond to one of the 
fourteen partitions of the external momenta. The non-vetoed 
graphs correspond to different one-loop diagrams: $a_1$, $a_4$ and 
$a_5$ are cuts contributing to three different triangles, whereas 
$a_2$ and $a_3$ contribute to two different box diagrams. Considering 
the whole set of fourteen partitions, one recovers the full 
four-point Green function. 
}
\end{figure}
\end{center}
\section{Numerical results}
\label{numerical}
We give, in this section, few examples of results obtained 
with the above described numerical method compared to standard analytic 
evaluation of loop integrals~\cite{pv}. More complete numerical results will 
be shown elsewhere. First of all we give the predictions 
in Tabs.~\ref{tab:scalarc},~\ref{tab:scalard},~\ref{tab:scalare}
for the three-, four- and five-points scalar form factors respectively, 
defined in Eq.~\ref{eq:formf}, in the $\phi^3$ scalar theory. 
We compare only the real part of the form factors, since the method 
illustrated in the previous sections allows to calculate it. 
The imaginary part can be easily calculated with standard 
Cutkowsky rules. This is much easier than the real part, 
since the required integration over the internal cut lines 
is only the 2-body phase-space and no dispersion integral is required. 
Moreover, no principal value integration is needed\footnote{ 
For the real part of the amplitude we deal with the numerical evaluation 
of  principal value integrals using the following strategy. 
For simplicity let's assume we have to
 perform the principal value integral of a function
$f(x)$ with a pole in $x_0$. We can either
perform the integral of
\[
f_1(x)=\frac{1}{2} [f(x)+f(2 x_0-x)]
\]
or, use integration by part,
\begin{eqnarray}
\int f(x) \frac{x-x_0}{x-x_0} \mathrm d x && \to \log |x-x_0| 
(x-x_0) f(x) \nonumber \\ 
&& - \int  \log |x-x_0|
\frac{\mathrm d }{\mathrm d x} [(x-x_0) f(x)] \mathrm d x \nonumber
\end{eqnarray}
in both cases obtaining a {\em numerically} convergent 
integral. Both methods have been explored and tested. 
(We however need to study computationally
more involved examples to be sure that numerical instabilities do not 
arise because of nearby poles or soft/collinear enhancements).
Notice that we can apply the method here sketched since we can easily evaluate
analytically the location of the poles as a function of the integration variables. }.

\begin{eqnarray}
C_0 &=& \frac{1}{i \pi^2}\int d^4q 
\frac{1}{d_1 d_2 d_3} ,   \nonumber \\
D_0 &=& \frac{1}{i \pi^2}\int d^4q \frac{1}{d_1 d_2 d_3 d_4} ,\nonumber \\
E_0 &=& \frac{1}{i \pi^2}\int d^4q \frac{1}{d_1 d_2 d_3 d_4 d_5} , 
\label{eq:formf}
\end{eqnarray}
where
\begin{eqnarray}
d_1 &=& (q^2-m^2+i\epsilon)\nonumber\\
d_i&=&((q+k_{i-1})^2-m^2+i\epsilon)\nonumber
\end{eqnarray}
and $k_i$ are related to the external momenta as 
$p_1=k_1$, $p_n=k_n-k_{(n-1)}$.

\begin{table}
\begin{center}
\begin{tabular}{|c|c|c|c|}
\hline
         & $p_x$ (GeV) & $p_y$ (GeV) & $p_z$ (GeV) \\
\hline
$p_1$    &    $1.73205081 $        &    $1.41421354$        & $-2.23606801$ \\
$p_2$    &    $-1.73205081 $        &    $-1.41421354 $        & $2.23606801 $ \\
$p_3$    &    $0$        &    $0$         & $0$ \\
\hline
 & Re$(C_0)$      &   1.982(2) (numerical)   &  
1.98390995 ~Ref.~\cite{looptools}\\        
\hline
\end{tabular}
\caption{\label{tab:scalarc} Comparison between the 
values of the scalar form factor $C_0$ for the $\phi^3$ theory 
as obtained with the numerical method and with the analytical 
results of LoopTools~\cite{looptools}. The momenta components are 
taken to be $p=(p_x,p_y,p_z)$, with a common internal and external 
mass of 0.01~GeV.} 
\end{center}
\end{table}

\begin{table}
\begin{center}
\begin{tabular}{|c|c|c|c|}
\hline
         & $p_x$ (GeV) & $p_y$ (GeV) & $p_z$ (GeV) \\
\hline
$p_1$    &    $1.73205081 $        &    $1.41421354$        & $-2.23606801$ \\
$p_2$    &    $-1.73205081 $        &    $-1.41421354 $        & $2.23606801 $ \\
$p_3$    &    $-2.23606801 $        &    $-1.41421354 $         & $-1.73205081 $ \\
$p_4$    &    $2.23606801 $        &    $1.41421354 $         & $1.73205081$ \\
\hline
   & Re$(D_0)$       &  0.4435(2) (numerical)  &  
0.443615191 ~Ref.~\cite{looptools}  \\
\hline
\end{tabular}
\caption{\label{tab:scalard} Comparison between the 
values of the scalar form factor $D_0$ for the $\phi^3$ theory 
as obtained with the numerical method and with the analytical 
results of LoopTools~\cite{looptools}. The momenta components are 
taken to be $p=(p_x,p_y,p_z)$, with a common internal and external 
mass of 0.01~GeV.} 
\end{center}
\end{table}

\begin{table}
\begin{center}
\begin{tabular}{|c|c|c|c|}
\hline
            & $p_x$ (GeV) & $p_y$ (GeV) & $p_z$ (GeV) \\
\hline
$p_1$       &    $0$        &    $0$        & $2.13816766718398$ \\
$p_2$       &    $0$        &    $0$        & $-2.13816766718398$ \\
$p_3$  &$-0.825139760971069$& $-0.878521561622620$ &$-0.08667117357254028$ \\
$p_4$  &$-0.501121819019318$& $-0.772821187973022$ &$-0.162467777729034$   \\
$p_5$  &$1.32626157999039$&   $1.65134274959564$   &$0.249138951301575$ \\   
\hline
       &       Re($E_0$)     &  -10.48(2) (numerical)  &  
-10.4724461~Ref.~\cite{looptools}          \\
\hline
\end{tabular}
\caption{\label{tab:scalare}  Comparison between the 
values of the scalar form factors for the $\phi^3$ theory 
as obtained with the numerical method and with the analytical 
results of LoopTools~\cite{looptools}. The momenta components are 
taken to be $p=(p_x,p_y,p_z)$, with a common internal and external 
mass of 0.01~GeV.} 
\end{center}
\end{table}

In addition we compare the pure virtual NLO QCD correction to 
$e^+ e^- \to q \bar q$ production (limited to the photon exchange case), 
at $\sqrt{s}=20$~GeV and for different values of the fermion mass. 
The infrared singularity has been regulated giving a 
mass of 30~MeV to the gluon. Since no QCD non-abelian vertex is 
present the gluon mass can be safely used. 
The QCD coupling $g_s$ has been set to 1 for simplicity. 

For $e^+e^-\to q\bar q$ there is a divergent 3-point function.
We deal with the renormalization problem as follows:
\begin{itemize}
\item We compute the 3-point function using a hard cut-off $\tau_{\rm max}$.
\item For a specific external momenta configuration we compute the 3-point Green function both using the hard cut-off $\tau_{\rm max}$, obtaining $G_3^{(\tau_{\rm max})}$, and with a conventional regularization, e.g. dimensional regularization (DR), obtaining $G_3^{(DR)}$.
\item We then add to the Lagrangian the counterterm\newline $G_3^{(DR)}-G_3^{(\tau_{\rm max})}=\bar\psi(a_0^{(\tau_{\rm max})}\gamma_0 A_0-a^{(\tau_{\rm max})}{\bf \gamma}\cdot{\bf A})\psi$.
\end{itemize}

\begin{table}
\begin{center}
\begin{tabular}{|c|c|c|}
\hline
fermion mass (GeV) & analytical~(nb) & numerical~(nb) \\
\hline
1         & -1.0622(2)          &  -1.0623(3)           \\
2         & -0.7879(1)          &  -0.7882(3)           \\
3         & -0.5985(1)          &  -0.5989(2)           \\
8         & +0.01320(0)         &  +0.01289(8)          \\
9         & +0.10271(0)         &  +0.10258(6)          \\
\hline
\end{tabular}
\caption{\label{tab:xsect} Values of the virtual QCD cross section 
for the process $e^+ e^- \to \gamma^* \to q \bar q$, at $\sqrt{s} = 20$~GeV, 
for different fermion masses. The infrared divergence has been regulated 
with a photon mass $m_\gamma = 30$~MeV.}
\end{center}
\end{table}

\section{Conclusions}
We have proposed a new approach 
to the evaluation of one-loop scattering 
amplitudes in perturbative quantum field theory. 
In the present paper we have presented the 
method via some basic examples of one-loop amplitudes 
in $\phi^3_4$ field theory. The basic 
rules of our calculation approach have been briefly 
illustrated. 

We have calculated few 
scalar Green functions and evaluated the 
QCD next-to-leading order correction to $e^+e^- \to \gamma^* \to q \bar q$ 
finding very good agreement with analytical results.

To assess the viability of the approach there are still 
several steps to be faced. We plan to: 

\bit

\item Apply the method to the evaluation of n-photon 
amplitudes in QED.

\item Explore the feasibility of a regularization based on a 
hard momentum cut-off both in the ultraviolet and in the infrared region
for non abelian gauge theories. 
This should be possible 
provided a suitable number of counterterms
is introduced in the Lagrangian~\cite{bdm95}.

\item Reproduce available results in $e^+e^- \to $jets 
at next-to-leading order accuracy.

\item Establish the relationship among our regularization
scheme and the standard dimensional regularization in order
to be able to use existent parton distribution 
functions. 

\eit

\vskip 8pt\noindent
{\bf Acknowledgments} \\ 
We are grateful to S. Catani and G. Marchesini 
for useful discussions. 

\section{Appendix A}

\begin{center}
\begin{figure}[hbt]
\SetScale{0.5}
\SetWidth{1.}
{\unitlength 0.5pt
\begin{picture}(0,0)(-100,100)
\ArrowLine(150,0)(300,0)
\CArc(150,0)(7,0,360)
\put(50,-5){$\Delta_{ij}^+$}
\put(400,-5){$x_i^0 > x_j^0$}
\ArrowLine(150,-100)(300,-100)
\CArc(300,-100)(7,0,360)
\put(50,-105){$\Delta_{ij}^-$}
\put(400,-105){$x_j^0 > x_i^0$}
\ArrowLine(150,-200)(300,-200)
\CArc(150,-200)(7,0,360)
\CArc(300,-200)(7,0,360)
\put(50,-210){$\Delta_{ij}^*$}
\end{picture}
}
\vskip 180pt
\ccaption{}{\label{feynbull}
Feynman rules for propagators involving at least one dotted vertex}
\end{figure}
\end{center}
For the self energy diagram we have (see Eqn.(\ref{prpg}))
\begin{eqnarray}
G(x,y)= \Delta_{xy} \Delta_{yx} 
           &=& (\vartheta_{xy} \Delta^+_{xy}
                + \vartheta_{yx} \Delta^-_{yx})
               (\vartheta_{yx} \Delta^+_{yx}
                + \vartheta_{xy} \Delta^-_{yx}) \nonumber \\
           &=& \vartheta_{xy} (\Delta^+_{xy})^2 
                + \vartheta_{yx} (\Delta^+_{yx})^2 \nonumber \\
           &=& \vartheta_{xy} \left\{ \frac{1}{(2 \pi)^6} \int \dr^4 k\, \
\dr^4 q\, 
               \exp[-i(x-y)(k+q)] \right. \nonumber \\
            & & \left. \quad \quad \quad \times \vartheta(k^0) \vartheta(q^0) 
               \delta(k^2 - m_1^2) \delta(q^2 - m_2^2) \right\} \nonumber \\
           &+& \vartheta_{yx} \left\{ \frac{1}{(2 \pi)^6} \int \dr^4 k\, 
\dr^4 q\,  
               \exp[-i(y-x)(k+q)] \right. \nonumber \\ 
            & & \left. \quad \quad \quad \times \vartheta(k^0) \vartheta(q^0) 
               \delta(k^2 - m_1^2) \delta(q^2 - m_2^2) \right\},
\label {slfstnd}
\eeanmb
where $\vartheta_{yx} =\theta(x_0-y_0)$.
We can now introduce an integral representation for the $\theta$
function
\[
\theta(x)= \frac{1}{2 \pi i}
\int \frac {\er^{i \tau x}} {\tau-i \epsilon} \ \dr \tau \,, 
\]
with $\epsilon \to 0^+$ and Eqn.~(\ref{slfstnd}) becomes

\beanmb
G(x,y)
           &=& \frac{1}{2 \pi i} \int 
            d \tau \frac{\exp[i \tau (x^0 - y^0)]}{\tau-i\epsilon}
 \nonumber \\ 
           &\times& \left\{ \frac{1}{(2 \pi)^6} \int \dr^4 k\, \dr^4 q\,  
               \exp[-i(x-y)(k+q)] \right. \nonumber \\
           && \quad \quad \quad \left. \times \vartheta(k^0) \vartheta(q^0) 
               \delta(k^2 - m_1^2) \delta(q^2 - m_2^2) \right\} \nonumber \\
           &+& \frac{1}{2 \pi i} \int
           d \tau \frac{\exp[-i \tau (x^0 - y^0)]}{\tau -i \epsilon}
 \nonumber \\ 
           &\times& \left\{ \frac{1}{(2 \pi)^6} \int \dr^3 k\, \dr^3 q\, 
               \exp[-i(y-x)(k+q)] \right. \nonumber \\ 
           && \quad \quad \quad \left. \times \vartheta(k^0) \vartheta(q^0) 
               \delta(k^2 - m_1^2) \delta(q^2 - m_2^2) \right\} \nonumber 
\eeanmb

Since we want to compute the self energy in momentum space, we perform
the Fourier transform 
\[
G(p_1,p_2) = \int \er^{i p_1 x} \er^{i p_2 y} G(x,y) \ \dr x \ \dr y
{\buildrel {w=x+y} \over {=}}
\int \er^{i p_1 x} \er^{i p_2 (w + x)} G(w)\ \dr x \ \dr w
\]
and we obtain
\begin{eqnarray}
G(p_1,p_2) &=& \frac{1}{(2 \pi)^3 i} (2\pi)^4\delta(p_1+p_2)
\nonumber \\
& & \int \dr \tau \int  \dr^4 k \ \dr^4 q 
\frac{1}{\tau-i \epsilon} \delta^3(p_2+k+q)
\vartheta(k^0) \vartheta(q^0) \delta(k^2 - m_1^2) \delta(q^2 - m_2^2) 
\nonumber \\&   \times &
[ \delta (\tau +p_{20}+k_0+q_0) + \delta (\tau -p_{20}+k_0+q_0) ]
\nonumber \\ 
& = & \frac{1}{(2 \pi)^3 i} (2\pi)^4\delta(p_1+p_2) 
\int \dr \tau \frac{1}{\tau-i \epsilon}
\int  \dr^3 k \frac{1}{2 k_0}  \frac{1}{2 q_0}
\nonumber \\&   \times &
[ \delta (\tau +p_{20}+k_0+q_0) + \delta (\tau -p_{20}+k_0+q_0) ] 
\end{eqnarray}
where, to perform $x$ and $w$ integrations,
 we have used $\int \er^{i k x} \dr^4 x =(2 \pi)^4 \delta^4(k)$
and $k_0=\sqrt{k^2+m_1^2}$, $q_0=\sqrt{(p_2+k)^2+m_2^2}$.
Recalling that
\[
\int \frac {f(\tau)}{\tau-i\epsilon}\ \dr \tau  
= \fint \frac {f(\tau)}{\tau}\ \dr \tau + i \pi \int f(\tau) \delta(\tau)  \ \dr \tau
\]
we obtain
\begin{eqnarray}
G(p_1,p_2) 
& = & \frac{1}{(2 \pi)^3 i} (2\pi)^4\delta(p_1+p_2) 
\left \{
\fint \dr \tau \frac{1}{\tau}
\int  \dr^3 k \frac{1}{2 k_0}  \frac{1}{2 q_0} \right .
\nonumber \\&   \times &
[ \delta (\tau +p_{20}+k_0+q_0) + \delta (\tau -p_{20}+k_0+q_0) ] 
\nonumber \\ 
& & \left .
+ i \pi \int  \dr^3 k \frac{1}{2 k_0}  \frac{1}{2 q_0} 
\delta (p_{20}-k_0-q_0) \right \} \label{slfct}
\end{eqnarray}
where we assume $p_{20}>0$ and thus $\delta (p_{20}+k_0+q_0)=0$. We can
work in the $p_2$ rest-frame where 
 $k_0=\sqrt{k^2+m_1^2}$, $q_0=\sqrt{k^2+m_2^2}$.
The phase space volume $\dr^3 k$ can be written as
$k^2 \dr k \ \dr \Omega$ and the integral over $\dr k$ can 
be performed eliminating the $\delta$ functions in Eqn.~(\ref{slfct})
and we finally obtain
\begin{eqnarray}
G(p_1,p_2) 
& = & \frac{1}{(2 \pi)^3 i} (2\pi)^4\delta(p_1+p_2) 
\left \{
\fint_{-\infty}^{-m_1-m_2-p_{20}}
 \dr \tau \frac{1}{\tau} \int  \dr \Omega \frac{k_+}{-\tau - p_{20}}
\right . 
\nonumber \\
& & \left .
+
\fint_{-\infty}^{-m_1-m_2+p_{20}} \dr \tau \frac{1}{\tau} \int  \dr \Omega \frac{k_-}{-\tau + p_{20}}+ i \pi \int \dr \Omega  \frac{k_i}{ p_{20}}
 \right \} 
\nonumber \\
k^2_\pm & = & 
\frac{(p_{20} \pm \tau)^2}{4} - \frac{m_1^2 + m_2^2}{2} 
+ \frac{(m_1^2 - m_2^2)^2}{4 (p_{20} \pm \tau)^2}
\nonumber \\
k^2_i & = & 
\frac{p_0^2}{4} - \frac{m_1^2 + m_2^2}{2} 
+ \frac{(m_1^2 - m_2^2)^2}{4 p_{20}^2}
\nonumber
\eeanmb
The last term correspond to the real part of the diagram and is indeed equal to
half the contribution found, in Eqn.~(\ref{rlslf}), using the largest
time equation. The first two terms correspond to the imaginary
part of the integrals and coincide with the result of Eqn.~(\ref{imslf}).
Notice that the dispersive integral over $\dr \tau$ arises as a consequence
of using the integral repesentation for the $\theta$ function, thus in 
the equation (\ref{eq:realpart2pt}) 
\[
[\theta(x_0-y_0)-\theta(y_0-x_0) ] \to \frac{1}{i\pi} \fint \frac{1}{\tau}
\er^{i \tau (x_0-y_0)}
\]
and it is responsible for the appearence of the dispersive integral
as well as of the appearence of $\tau$ into energy momentum conservation
$\delta$ functions.

\section{Appendix B}
We shall now explicitly verify Eqn.~(\ref{eq:lgst3pt}).
We have
\beanmb
(1/i^3) \tilde G(x,y,z) &=& i^3 \Delta_{xy}
\Delta_{yz} \Delta_{zx}
 =  \theta_{xy}\theta_{yz}\Delta^+_{xy}\Delta^+_{yz}\Delta^+_{xz}
+\theta_{xy}\theta_{zx}\Delta^+_{xy}\Delta^+_{zy}\Delta^+_{zx}
\nonumber \\
& & +\theta_{zy}\theta_{xz}\Delta^+_{xy}\Delta^+_{zy}\Delta^+_{xz}
+\theta_{yx}\theta_{xz}\Delta^+_{yx}\Delta^+_{yz}\Delta^+_{xz}
\nonumber \\
& & +\theta_{yz}\theta_{zx}\Delta^+_{yx}\Delta^+_{yz}\Delta^+_{zx}
+\theta_{yx}\theta_{zy}\Delta^+_{yx}\Delta^+_{zy}\Delta^+_{zx}
\nonumber \\
& = & \theta_{xy}\theta_{yz}\Delta^+_{xy}\Delta^+_{yz}\Delta^+_{xz}
+\theta_{xy}\theta_{zx}\Delta^+_{xy}\Delta^+_{zy}\Delta^+_{zx}
\nonumber \\
& &+\Delta^+_{xy}\Delta^+_{zy}\Delta^+_{xz}
 -\theta_{yz}\Delta^+_{xy}\Delta^+_{zy}\Delta^+_{xz} \nonumber \\ 
& &-\theta_{zx}\Delta^+_{xy}\Delta^+_{zy}\Delta^+_{xz} 
+\theta_{yz}\theta_{zx}\Delta^+_{xy}\Delta^+_{zy}\Delta^+_{xz}
\nonumber \\ 
& & +\Delta^+_{yx}\Delta^+_{yz}\Delta^+_{xz}
-\theta_{xy}\Delta^+_{yx}\Delta^+_{yz}\Delta^+_{xz} \nonumber \\
& &-\theta_{zx}\Delta^+_{yx}\Delta^+_{yz}\Delta^+_{xz}
+\theta_{xy}\theta_{zx}\Delta^+_{yx}\Delta^+_{yz}\Delta^+_{xz} \nonumber \\
& &+\theta_{yz}\theta_{zx}\Delta^+_{yx}\Delta^+_{yz}\Delta^+_{zx}
+\Delta^+_{yx}\Delta^+_{zy}\Delta^+_{zx}
-\theta_{xy}\Delta^+_{yx}\Delta^+_{zy}\Delta^+_{zx} \nonumber \\
& &-\theta_{yz}\Delta^+_{yx}\Delta^+_{zy}\Delta^+_{zx}
+\theta_{xy}\theta_{yz}\Delta^+_{yx}\Delta^+_{zy}\Delta^+_{zx}
\nonumber \\
& = &
\Delta^+_{xy}\Delta^+_{zy}\Delta^+_{xz}
+\Delta^+_{yx}\Delta^+_{yz}\Delta^+_{xz}
+\Delta^+_{yx}\Delta^+_{zy}\Delta^+_{zx} \nonumber \\
& &-\theta_{yz}\left (\Delta^+_{xy}\Delta^+_{zy}\Delta^+_{xz} 
 + \Delta^+_{yx}\Delta^+_{zy}\Delta^+_{zx}
\right ) \nonumber \\ 
& & 
-\theta_{zx}
\left ( \Delta^+_{xy}\Delta^+_{zy}\Delta^+_{xz}
+
\Delta^+_{yx}\Delta^+_{yz}\Delta^+_{xz}
\right ) \nonumber \\
& &
-\theta_{xy} \left (
\Delta^+_{yx}\Delta^+_{yz}\Delta^+_{xz}
+
\Delta^+_{yx}\Delta^+_{zy}\Delta^+_{zx}
\right ) \nonumber \\ 
& & 
+\theta_{xy}\theta_{yz} \left (
\Delta^+_{xy}\Delta^+_{yz}\Delta^+_{xz}+
\Delta^+_{yx}\Delta^+_{zy}\Delta^+_{zx}
\right ) \nonumber \\
& & +
\theta_{xy}\theta_{zx}
\left ( \Delta^+_{xy}\Delta^+_{zy}\Delta^+_{zx}
+
\Delta^+_{yx}\Delta^+_{yz}\Delta^+_{xz} \right )
\nonumber \\ 
& & 
+ \theta_{yz}\theta_{zx} \left (
\Delta^+_{xy}\Delta^+_{zy}\Delta^+_{xz}
+
\Delta^+_{yx}\Delta^+_{yz}\Delta^+_{zx}
\right )
\nonumber 
\label{appb1}
\eeanmb
where
$\theta_{AB}=\theta(A_0-B_0)$ and we have
repeatedly used 
$\theta_{AB}=1-\theta_{BA}$ 

Analogously we have
\beanmb
(1/i^3) \tilde G (\underline x ,y ,z)
& = & \theta_{yz}
\left( \Delta^+_{yx}\Delta^+_{zx}\Delta^+_{zy}
- \Delta^+_{yx}\Delta^+_{zx}\Delta^+_{yz}
\right )
- \Delta^+_{yx}\Delta^+_{zx}\Delta^+_{zy} 
\nonumber \nonumber \\
(1/i^3) \tilde G (x, \underline y ,z)
& = & \theta_{zx}
\left( \Delta^+_{xy}\Delta^+_{zy}\Delta^+_{xz}
- \Delta^+_{xy}\Delta^+_{zy}\Delta^+_{zx}
\right )
- \Delta^+_{xy}\Delta^+_{zy}\Delta^+_{xz} \nonumber \\
\nonumber \nonumber \\
(1/i^3) \tilde G (x, y, \underline z)
& = & \theta_{xy}
\left( \Delta^+_{xz}\Delta^+_{yz}\Delta^+_{yx}
- \Delta^+_{xz}\Delta^+_{yz}\Delta^+_{xy}
\right )
- \Delta^+_{xz}\Delta^+_{yz}\Delta^+_{yx} \nonumber \\
(1/i^3) \tilde G (x, \underline y, \underline z)
& = & \theta_{yz}
\left( \Delta^+_{xy}\Delta^+_{xz}\Delta^+_{yz}
- \Delta^+_{xy}\Delta^+_{xz}\Delta^+_{zy}
\right )
+ \Delta^+_{xy}\Delta^+_{xz}\Delta^+_{zy} \nonumber \\
(1/i^3) \tilde G (\underline x,  y, \underline z)
& = & \theta_{xz}
\left( \Delta^+_{yx}\Delta^+_{zy}\Delta^+_{xz}
- \Delta^+_{yx}\Delta^+_{zy}\Delta^+_{zx}
\right )
+ \Delta^+_{yx}\Delta^+_{zy}\Delta^+_{zx} \nonumber \\
(1/i^3) \tilde G (\underline x,  \underline y, z)
& = & \theta_{xy}
\left( \Delta^+_{xz}\Delta^+_{yz}\Delta^+_{xy}
- \Delta^+_{xz}\Delta^+_{yz}\Delta^+_{yx}
\right )
+ \Delta^+_{xz}\Delta^+_{yz}\Delta^+_{yx} \nonumber \\
(1/i^3) \tilde G (\underline x,  \underline y, \underline
z) & = &
\Delta^+_{yx}\Delta^+_{yz}\Delta^+_{zx}
-
\Delta^+_{xy}\Delta^+_{zy}\Delta^+_{zx}
-
\Delta^+_{xy}\Delta^+_{yz}\Delta^+_{xz} \nonumber \\
& & +
\theta_{yz}
\left (
\Delta^+_{yx}\Delta^+_{yz}\Delta^+_{zx}
+
\Delta^+_{xy}\Delta^+_{yz}\Delta^+_{xz}
\right )
\nonumber \\ 
& & +
\theta_{zx}
\left ( \Delta^+_{yx}\Delta^+_{yz}\Delta^+_{zx}
+
\Delta^+_{xy}\Delta^+_{zy}\Delta^+_{zx}
\right ) \nonumber \\
& & +
\theta_{xy} \left (
\Delta^+_{xy}\Delta^+_{zy}\Delta^+_{zx}
+
\Delta^+_{xy}\Delta^+_{yz}\Delta^+_{xz}
\right ) \nonumber \\ 
& & 
-
\theta_{xy}\theta_{yz} \left (
\Delta^+_{xy}\Delta^+_{yz}\Delta^+_{xz}+
\Delta^+_{yx}\Delta^+_{zy}\Delta^+_{zx}
\right ) \nonumber \\
& & 
-
\theta_{xy}\theta_{zx}
\left ( \Delta^+_{xy}\Delta^+_{zy}\Delta^+_{zx}
+
\Delta^+_{yx}\Delta^+_{yz}\Delta^+_{xz} \right )
\nonumber \\ & & 
- \theta_{yz}\theta_{zx} \left (
\Delta^+_{xy}\Delta^+_{zy}\Delta^+_{xz}
+
\Delta^+_{yx}\Delta^+_{yz}\Delta^+_{zx}
\right )
\label{appb2}
\eeanmb
and from Eqns.~(\ref{appb1},\ref{appb2}) we 
indeed obtain Eqn.(\ref{eq:lgst3pt}).

\end{document}